\documentclass[aps,pre,reprint,]{revtex4-2}

\usepackage{amsmath}
\usepackage{amssymb}
\usepackage{graphicx}
\usepackage{dcolumn}
\usepackage{bm}
\usepackage{multirow}
\usepackage{makecell}
\usepackage{booktabs}
\usepackage{array}

\usepackage[T1]{fontenc}
\graphicspath{{Figures/}}

\begin{document}

\title{Ferri- and Ferro-Electric Switching in Spontaneously Chiral Polar Liquid Crystals}

\author{Jordan Hobbs$^1$}
\email{j.l.hobbs@leeds.ac.uk}
\author{Calum J. Gibb$^2$}
\author{Richard. J. Mandle$^{1,2}$}
\affiliation{$^1$School of Physics \& Astronomy, University of Leeds, UK}
\affiliation{$^2$School of Chemistry, University of Leeds, UK}

\date{\today}

\begin{abstract}
The recent discovery of spontaneous chiral symmetry breaking has demonstrated the possibility of discovering the exotic textures of ferromagnetic systems in liquid crystalline fluid ferro-electrics. We show that the polar smectic mesophase exhibited by the first molecule discovered to exhibit a spontaneously chiral ferroelectric nematic phase is also helical has a strongly varied textural morphology depending in its thermal history and phase ordering. Electro-optic studies demonstrate that the two spontaneously chiral phases exhibit field induced phase transitions. For the nematic variant, this process is threshold-less and has no hysteresis while for the smectic it has a clear threshold and shows hysteresis meaning this phase exhibits pseudo-ferrielectric switching, the first of its kind for ferroelectric nematic like phases. We show that helix formation can be both 1st and 2nd order but when it is 1st it is accompanied by pre-transitional helix formation in the preceding ferroelectric nematic phase.
\end{abstract}

\maketitle

\section{Introduction}

Spontaneous symmetry breaking is a general phenomena where a symmetric system will undergo some transition to an asymmetric state where it is no longer invariant under some symmetry operation and has wide ranging consequences for the natural world such as homochirality in biology \cite{2010_Blackmond}, autocatalysis in chemistry \cite{2019_Shen}, the Higgs mechanism in physics \cite{1964_Higgs} and even in the understanding of crowd control \cite{2025_Gu}. Liquid crystals can have various components of order (orientational, translational, bond orientation etc.) and, as such, symmetry breaking can have profound effects on the properties exhibited by the phase \cite{2018_Tschierske}. Ferroelectric liquid crystals are notable examples of symmetry breaking within soft matter research resulting in a variety of phase structures with distinctly different properties. Chiral smectic C liquid crystals (SmC*) form a tilted helical structure where the molecular ordering of dipole moments associated with the chiral centre break mirror symmetry resulting in an orthogonal polar phase that can be made ferroelectric upon proper confinement \cite{2004_Lagerwall, 2010_Takezoe}. Bent core liquid crystals can spontaneously break inversion symmetry forming phase structures containing layer polarisations of various types \cite{2013_Eremin,2018_Jakli} as well as also spontaneously breaking chiral symmetry where the layer polarisation rotates around a helix \cite{2016_Sreenilayam}. The twist bend nematic (N$_{\text{TB}}$) is another example of chiral symmetry breaking, forming a phase that has degenerate left and right handed heliconic molecular organisation resulting in a phase that is both chiral and locally polar \cite{2018_Jakli, 2022_Imrie, 2022_Mandle}.

In 2017, the ferroelectric nematic phase (N$_{\text{F}}$) phase was discovered \cite{2017_Mandle_1, 2017_Mandle_2, 2017_Nishikawa} in which the molecules spontaneously align syn-parallel to each other thus breaking inversion symmetry and resulting in longitudinally polar phases \cite{2018_Mertelj, 2020_Chen, 2020_Sebastian, 2021_Mandle, 2021_Li, 2022_Sebastian, 2024_Kumari, 2024_Cruickshank, 2025_Gibb}. It has been shown that the N$_{\text{F}}$ phase can spontaneously break chiral symmetry to form $\pi$-twisted domains \cite{2021_Chen} to reduce the electrostatic cost of polarisation \cite{2024_Kumari}. Further investigations into novel N$_{\text{F}}$ materials led to the discovery of additional longitudinally polar phases \cite{2022_Chen2, 2022_Kikuchi, 2023_Chen, 2024_Gibb, 2024_Hobbs2, 2024_Kikuchi, 2024_Strachan, 2025_Rupnik} and two that showed not only broken inversion symmetry but also distinct spontaneously chiral heliconic molecular organisation \cite{2024_Karcz, 2024_Gibb}. 

The so called ferroelectric twist bend nematic (N$_{\text{TBF}}$) phase \cite{2024_Karcz, 2025_Strachan} (also referred to a $^{\text{HC}}$N$_{\text{F}}$ \cite{2024_Nishikawa}) and polar helical smectic C (SmC$_\text{P}^\text{H}$) phase \cite{2024_Gibb} both spontaneously form heliconic ordering of the LC molecules with a temperature dependent pitch around $\sim$400-1000 nm resulting in selective reflection of visible light. Notably, below the N$_{\text{TBF}}$ phase a further phase transition into a tilted smectic phase of unknown structure was observed \cite{2024_Karcz}. This raises the question: does the helix unwind in some fashion to form a non helical ferroelectric SmC phase, as has been recently discovered \cite{2024_Hobbs2, 2024_Kikuchi, 2024_Strachan}, or does the helix remain meaning the phase is actually a SmC$_\text{P}^\text{H}$ phase?

In this paper we demonstrate this to be the latter case, showing that this unknown smectic phase is a SmC$_\text{P}^\text{H}$ phase (we will refer to the SmC$_\text{P}^\text{H}$ phase in \textbf{2} briefly as SmC$_{\text{F}}$ throughout the first section of this article) showing continual existence of the heliconic structure and that the texture and defects of the SmC$_\text{P}^\text{H}$ phase varies strongly on the phase preceding it. We demonstrate that while both heliconic phases show a field induced phase transition into their non-helical and non-tilted counterparts, the SmC$_\text{P}^\text{H}$ exhibits hysteresis of this behaviour indicating ferrielectric switching while the N$_{\text{TBF}}$ does not. We also show that further transitions in the SmC$_\text{P}^\text{H}$ phase are possible and that these are accompanied by pre-transitional helix formation. 

\section{Materials}

\begin{table*}[]
\centering
\caption{Transition temperatures (T) and associated enthalpies of transition ($\Delta$H) for compounds 1–4 determined by DSC on cooling at a rate of 10 °C min$^{-1}$.}
\label{tbl:structures}
\resizebox{\textwidth}{!}{%
\begin{tabular}{ccccccccccccc}\toprule
 \multicolumn{12}{c}{\includegraphics[width=0.55\linewidth]{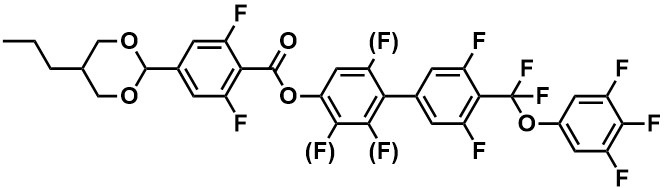}} \\ \midrule
 No. & & &
  \textbf{Melt} &
  \textbf{SmC$_\text{P}^\text{H}$-SmA$_\text{F}$} &
  \textbf{SmA$_\text{F}$-SmA} &
  \textbf{SmA-N} &
  \textbf{SmC$_\text{P}^\text{H}$-N$_\text{F}$} &
  \textbf{SmC$_\text{P}^\text{H}$-N$_\text{TBF}$} &
  \textbf{N$_\text{TBF}$-N$_\text{F}$} &
  \textbf{N$_\text{F}$-N$_\text{S}$} &
  \textbf{N$_\text{S}$-N} &
  \textbf{N-Iso} \\ \midrule
\textbf{1}& \raisebox{-.5\height}{\includegraphics[width=0.13\linewidth]{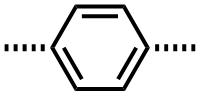}} & \makecell{T (°C)\\$\Delta$H (kJ/mol)} & \makecell{100.9\\31.1} & \makecell{90.1$^a$\\-} & \makecell{129.7\\0.6} & \makecell{154.3\\0.1} & - & - & - & - & - & \makecell{225.6\\1.0} \\
\textbf{2}& \raisebox{-.33\height}{\includegraphics[width=0.13\linewidth]{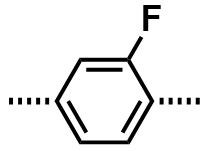}} & \makecell{T (°C)\\$\Delta$H (kJ/mol)} & \makecell{64.0\\19.0} & - & - & - & - & \makecell{91.6\\0.1} & \makecell{104.4$^a$\\-} & \makecell{142.6\\0.4} & \makecell{146.4\\0.01} & \makecell{225.1\\1.1} \\
\textbf{3}& \raisebox{-.5\height}{\includegraphics[width=0.13\linewidth]{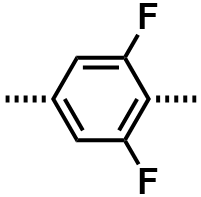}} & \makecell{T (°C)\\$\Delta$H (kJ/mol)} & \makecell{93.8\\19.2} & - & - & - & \makecell{90.6\\0.6} & - & - & \makecell{178.4\\0.7} & \makecell{179.5$^a$\\-} & \makecell{213.7\\1.3} \\
\textbf{4}& \raisebox{-.33\height}{\includegraphics[width=0.13\linewidth]{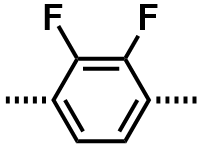}} & \makecell{T (°C)\\$\Delta$H (kJ/mol)} & \makecell{99.8\\25.1} & - & - & - & - & - & - & \makecell{119.3\\0.2} & \makecell{126.7\\0.01} & \makecell{233.8\\1.6} \\ \bottomrule
\multicolumn{12}{l}{Note: $^a$ determined via POM observations as no corresponding peak was observed in DSC}
\end{tabular}%
}
\end{table*}

Table \ref{tbl:structures} shows the chemical structures and phases transitions of the materials presented here. Details of  chemical synthesis and analysis (NMR, HPLC, HRMS) can be found in the ESI. All four compounds detailed here contain a 1,3-dioxane ring and so will show isomerisation from the equatorial trans state to the axial trans state just as for the classic ferroelectric mesogen DIO \cite{2024_Hobbs1}. While for DIO this isomerisation will occur above $\sim$140 °C the increased size of the molecules here increases the energy barrier for the isomerisation to occur. Our experiments suggest that quickly heating to $\sim$150 °C does not result in a reduction in transition temperatures but sustained time above this temperature results in a reduction in all transition temperatures indicative of isomerisation. Compounds \textbf{1} \cite{2024_Gibb} and \textbf{2} \cite{2024_Karcz} have been characterised previously and so their characterisation via DSC, X-ray, and Ps measurements are included in the ESI while compounds \textbf{3} and \textbf{4} will be discussed in section \ref{sec:2nd_materials}.

\section{Polarised Optical Microscopy} \label{sec:POM}

Compound \textbf{1} melts into the SmA$_\text{F}$ phase and in cells prepared with either parallel or anti-parallel rubbed surfaces forms blocky textures when viewed using polarised optical microscopy (POM) (\textbf{fig. S1a}). From azimuthal rotation of the cell with respect to the polarisers, it is clear that most of the blocky domains are orientated with the director close to the rubbing direction but with some slight in plane $\phi$ rotation diving the texture into domains. The blocky texture results from the restriction of splay deformations by the electrostatic cost of polarisation splay and the restriction of bend deformation by the formation of layers \cite{2022_Chen2, 2025_Hedlund}. Heating into the SmA phase retains these domains (\textbf{fig. S1b}) but they are lost upon reaching the N phase where a uniform planar texture is obtained. Upon cooling back into the SmA$_\text{F}$ phase a monodomain texture is obtained (\textbf{fig. S1c}) regardless of cell thickness. No evidence of director twist \cite{2021_Chen, 2024_Kumari} is found even in cells using parallel rubbed planar alignment layers.

\begin{figure*}
    \centering
    \includegraphics[width=1\linewidth]{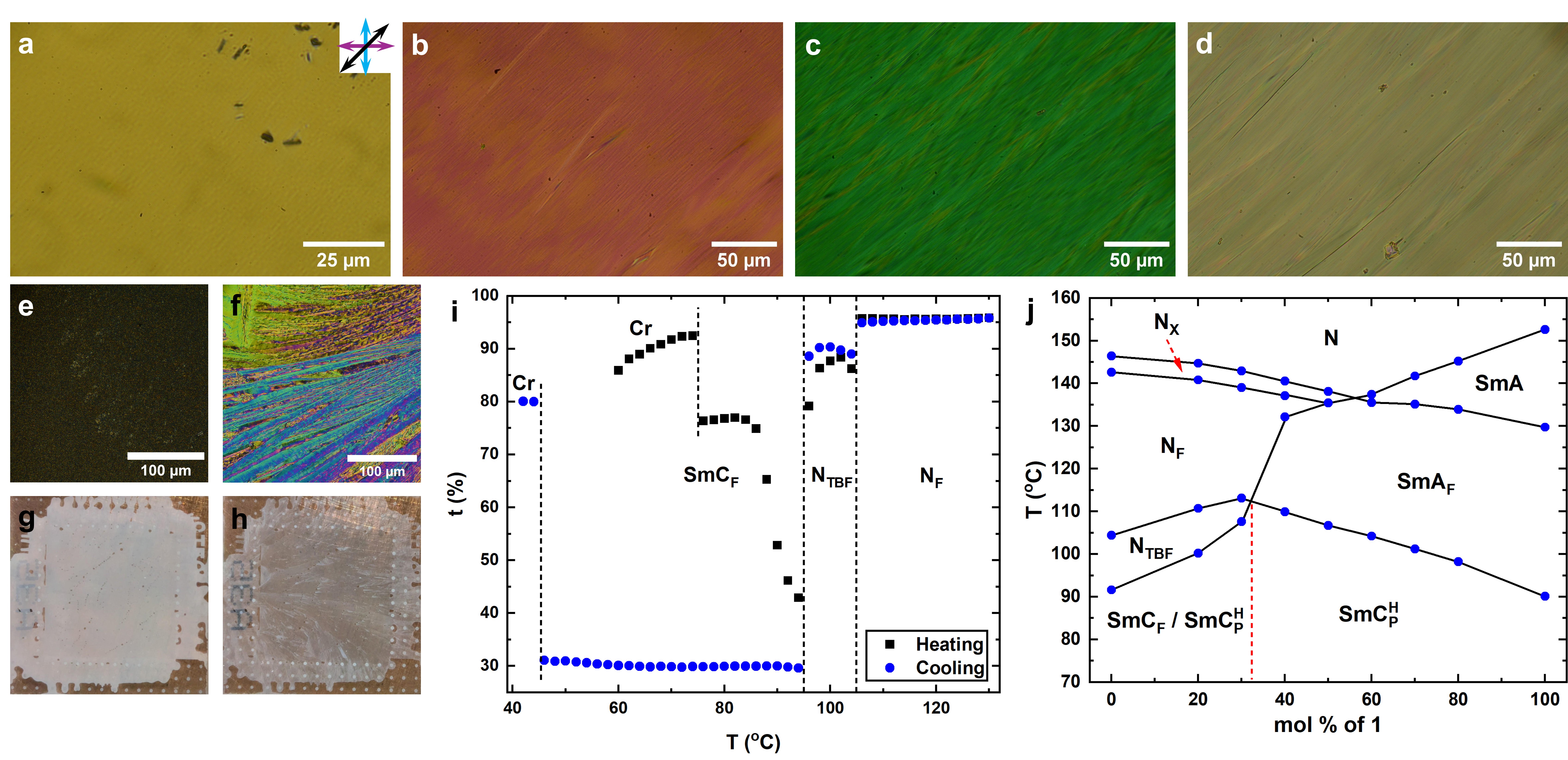}
    \caption{POM images taken for compound \textbf{1} at 80 °C filled into parallel rubbed, planar aligned cells with a cell gap of \textbf{a)} 1.6 $\mu m$, \textbf{b)} 5 $\mu m$, \textbf{c)} 10 $\mu m$ and \textbf{d)} 20 $\mu m$. POM images of \textbf{2} at 80 °C filled into parallel rubbed, planar aligned cell with a cell gap of 5 $\mu m$ on \textbf{e)} heating directly after melting from the crystal phase and \textbf{f)} subsequent cooling after heating to the nematic phase. Corresponding macroscopic texture of an entire 4 $\mu m$ cell filled with \textbf{2} at 80 °C on \textbf{g)} heating and \textbf{h)} cooling. The brown background colour comes from the hot stage plate. \textbf{i)} Transmission intensity of white light through a sample of \textbf{2} in a 5 $\mu m$ thick planar aligned parallel rubbed cell. \textbf{j)} Binary phase diagram between \textbf{1} and \textbf{2} with 100 \% compound 1 on the right and 100 \% compound 2 on the left. The red vertical dashed line marks the point where the SmC$_\text{P}^\text{H}$ phase switches polymorphism from the highly scattering to the striated as discussed in the text. For all POM images polarisor axis' and rubbing direction are marked by the blue, purple and black arrows respectively.}
    \label{fig:POM_Phase}
\end{figure*}

Immediately upon transitioning to the SmC$_\text{P}^\text{H}$ phase striations appear parallel to the rubbing direction with an average periodicity of around 1.2 $\mu m$ at 80 °C (\textbf{fig. \ref{fig:POM_Phase}}). While the average periodicity of these striations is generally independent of cell thickness, the periodicity takes a greater possible range of values for increasing cell thickness. Upon cooling these striations behave quite differently depending on cell thicknesses. For thin cells (1.6 $\mu m$) the striation periodicity decreases to around 0.8 $\mu m$ by crystallisation at around 55 °C (\textbf{fig. \ref{fig:POM_Phase}}a). For thicker cells (5 $\mu m$) the striations initially decrease in periodicity before larger less uniformly shaped striations grow in to become the dominant texture (\textbf{fig. \ref{fig:POM_Phase}}b). For 10 $\mu m$ thick cells the non-uniform striations grow into thick and distinct defect lines, though these do not have to encompass a full domain and do not show alternative optical activity upon slight de-crossing of the polarisers (\textbf{fig. \ref{fig:POM_Phase}}c). 20 $\mu m$ thick cells behave similarly to the 10 $\mu m$ thick cell except the temperature threshold for the defect lines is much closer to the SmA$_\text{F}$-SmC$_\text{P}^\text{H}$ transition (\textbf{fig. \ref{fig:POM_Phase}}d). These striations are clearly distinct from alternating polarisation domains \cite{2024_Gibb} and form even when a monodomain SmA$_\text{F}$ texture aligned under an in-plane electric field is cooled into the SmC$_\text{P}^\text{H}$ phase. They are also retained upon applying a weak in-plane electric field parallel to the polarisation direction.

The smectic phase, SmC$_\text{F}$, exhibited by \textbf{2} shows quite different behaviour where in cells of 1.6 $\mu m$, 5 $\mu m$, 10 $\mu m$ and 20 $\mu m$ thickness the texture is consistently grainy and nondescript (\textbf{fig. \ref{fig:POM_Phase}e}) when the sample is cooled from the preceding N$_\text{TBF}$ phase resulting in significant light scattering and low light transmission though the cell (\textbf{fig. \ref{fig:POM_Phase}i}). The SmC$_\text{F}$ phase in compound \textbf{2} is monotropic and the texture exhibited by the phase on melting directly from the crystal phase is distinctly different (\textbf{fig. \ref{fig:POM_Phase}f}). The melt texture is paramorphotic of the crystal phase but does not show any of the defects and scattering found in the phase on cooling. On heating the melted SmC$_\text{F}$ texture to $\sim$10 °C below the transition into the N$_\text{TBF}$ the defects found on cooling begin to form and the transmission through the sample drops significantly but does decrease to the values obtained on cooling. While selective reflection is difficult to be observed on cooling a faint blue colour can be seen in the SmC$_\text{F}$ phase with the naked eye on heating suggesting that this phase does retain the helical structure of the preceding N$_\text{TBF}$ phase.

\begin{figure*}[ht]
    \centering
    \includegraphics[width=1\linewidth]{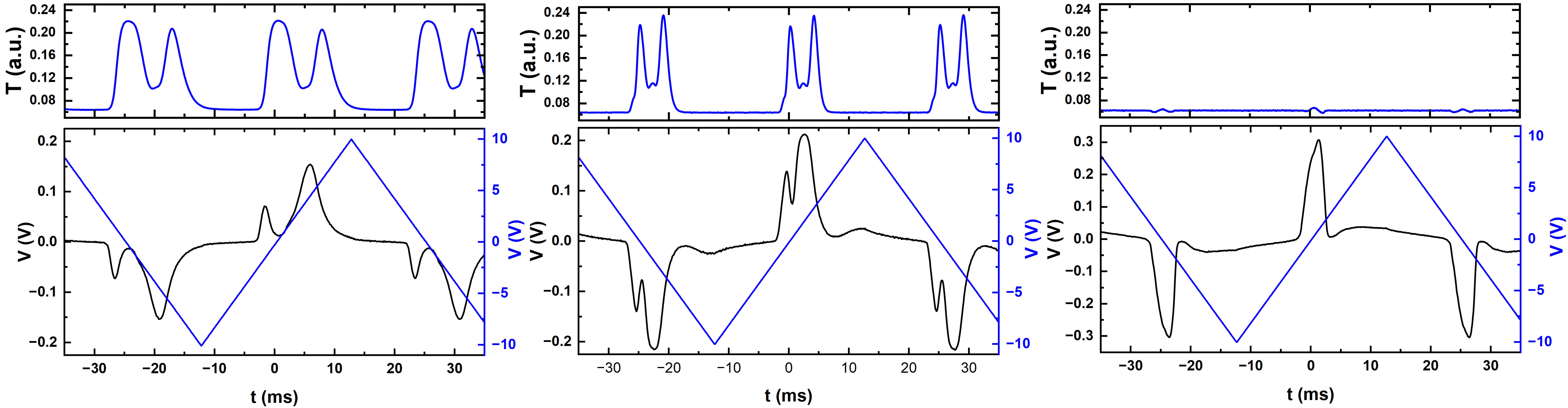}
    \caption{Electro-optic and current response to an 20 Hz, 10 V$_\text{RMS}$ applied triangle wave for \textbf{a)} compound 1 at 75°C, \textbf{b)} compound 2 at 75 °C and \textbf{c)} compound 2 at 99 °C. Each compound is filled into a 4 $\mu$m thick cell with bare ITO electrodes. The top portion of the graphs shows the transmission through the cell placed with between crossed polarisers as detected by a photo-diode while the bottom portion is the current response.}
    \label{fig:current_response}
\end{figure*}

If the SmC$_\text{F}$ phase is helical than the differences in optical textures on heating/cooling could result from the fact that the greatest change in pitch length (i.e. likely close to the phase transition) occurs in the same temperature regime as the scattering texture forms. These defects would then be the result of plastic deformation of the pitch though contraction and dilation \cite{1973_Clark}. However, so far the scattering texture have precluded optical measurement of any pitch length in the SmC$_\text{F}$ phase. We also note that the scattering behaviour of \textbf{2} in the SmC$_\text{F}$ phase is not isolated to cell confinement geometries and shows the same highly scattering and opaque appearance when filled into capillaries or as bulk samples in droplet form or in vials demonstrating that it is not a result of confinement.

\section{Binary Phase Diagram between \textbf{1} and \textbf{2}}

Construction of binary phase diagrams is a well established technique in liquid crystal science to demonstrate whether two materials posses the same phase by establishing miscibility between the two materials across the full phase diagram \cite{1983_Demus} and has been well applied to the N$_\text{F}$ phase already \cite{2022_Chen1}. Binary mixtures of \textbf{1} and \textbf{2} were created and the resulting phase diagram shown in \textbf{fig. \ref{fig:POM_Phase}j}. From the binary phase diagram of \textbf{1} and \textbf{2} the SmC$_\text{P}^\text{H}$ phase is completely miscible with the SmC$_\text{F}$ phase demonstrating that the SmC$_\text{F}$ phase is heliconic that the two phases are structurally the same. The studies presented so far have shown the SmC$_\text{F}$ found in \textbf{2} \cite{2024_Karcz} is the same as as the SmC$_\text{P}^\text{H}$ phase reported by us \cite{2024_Gibb} and so we shall refer to the SmC$_\text{F}$ phase of \textbf{2} as a SmC$_\text{P}^\text{H}$ phase from here on. 

The apolar-polar transition temperature (whether this is N-N$_\text{S}$ or SmA-SmA$_\text{F}$) is roughly linear with concentration while the transition to helical order, whether this is nematic of smectic type, is actually stabilised with a peak at around 30 \% of \textbf{1} for both the N$_\text{TBF}$ and the SmC$_\text{P}^\text{H}$ phase indicating that the molecular origins of ferroelectric order vs spontaneously chiral differ for these systems. It also points to the possibility of using mixtures in the future to construct systems with wide phase existence for application purposes. We note that the peak of chiral phase stability also roughly coincides with the point at which the transition pattern switches between N$_\text{F}$-N$_\text{TBF}$-SmC$_\text{P}^\text{H}$ to SmA$_\text{F}$-SmC$_\text{P}^\text{H}$. It is also this point where the texture and scattering behaviour changes from the highly scattering uncharacteristic texture to the striated texture. Seemingly, the key difference is the fact that the SmC$_\text{F}$ texture on cooling is from the N$_\text{TBF}$ phase while the SmC$_\text{P}^\text{H}$ texture is cooled from a SmA$_\text{F}$ phase.

The formation of the SmC$_\text{P}^\text{H}$ phase clearly differs depending on the phase preceding it and here we speculate on the origin of the difference in textures. There could be a significantly larger change in pitch at the N$_\text{TBF}$-SmC$_\text{F}$ transition vs the SmA$_\text{F}$-SmC$_\text{P}^\text{H}$, though this feels unlikely due to our expectation that the temperature dependence of the pitch at the SmA$_\text{F}$-SmC$_\text{P}^\text{H}$ transition will at least be somewhat similar to that at the N$_\text{F}$-N$_\text{TBF}$ transition i.e. a steep temperature dependence \cite{2024_Karcz}. We do note a very slight unwinding of the helix just before the N$_\text{TBF}$-SmC$_\text{F}$ transition (\textbf{fig. S13}). This is different from the much more significant unwinding of the helix observed for the transition from an N$_\text{TBF}$ phase to a non-helical SmC$_\text{P}$ phase \cite{2025_Strachan} where the helix unwinds critically.

An alternative is that forming layers in a system with an existing helical structure could result in significant defect formation due to potential deformation of the helix to form the regularly spaced smectic layers, although this does not seem to occur for the apolar twist bend nematic and associated smectic phases \cite{2019_Walker, 2021_Pociecha, 2024_Alshammari}, though the much smaller pitch, lack of polarity and vastly different origins (electric vs elastic) make this a less than optimal comparison.

Another alternative is that the mechanism of phase structure formation in the SmC$_\text{P}^\text{H}$ could vary significantly depending on the phase preceding it. This will mostly likely be true to some extent as observation of the temperature depndent tilt implies that the transition to the SmC$_\text{P}^\text{H}$ phase from the SmA$_\text{F}$ is only 2nd order while the N$_\text{TBF}$-SmC$_\text{F}$ transition is weakly first order, however whether this explains the differences in textures vs some alternative pitch defect effect remains to be determined. It could also be reminiscent to the formation of chevron defects in the SmC* phase where cooling from a SmA forms chevron structures due to surface pining \cite{1987_Rieker} which can be somewhat reduced when formed from a De Vries SmA phase where the randomly orientated molecular tilt reduces chevron formation \cite{2006_Lagerwall}.

\section{Electro-optics and Ferrielectric Switching}

\subsection{Current Response Measurements}

The current response of the SmC$_\text{P}^\text{H}$ of \textbf{1} in a 4 $\mu$m cell with out of plane bare ITO electrodes is not a single peak indicating complete reversal of the polarisation. Instead the current response shows a small peak pre-polarity reversal and a larger peak post-polarity reversal (\textbf{fig. \ref{fig:current_response}a}). As such in the original publication \cite{2024_Gibb} the phase was given the subscript designation of ``\textbf{P}'' to indicate polar, as the exact nature of the polarity was unclear from those initial measurements. We suggested that the small peak is associated with the emergence of molecular tilt. From the ground state at $E = 0$ V/$\mu$m increasing the electric field results in the first large peak corresponding to the full reversal of the polarisation vector from a SmC$_\text{P}^\text{H}$ state to a non-tilted, non-helical SmA$_\text{F}$ state. The small peak is then the reformation of the tilt and helix. The peak cannot be due to variation in the helical pitch length as this would not modulate the polarisation beyond a small variation due to changes in the region of partial pitch that may form due to the degenerate anchoring condition of the bare ITO electrode. Variation in the tilt angle, however, would modulate the effective polarisation observed upon switching. Throughout this next section we shall refer to the smaller peak as the ``tilt'' peak and the larger one as the ``polarisation reversal'' peak.

The following discussion is given in context of the current response corresponding to a negative to positive application of electric field. Perhaps unsurprisingly the SmC$_\text{P}^\text{H}$ phase of \textbf{2} shows a similar current response to that of \textbf{1} in terms of a tilt peak and a polarisation reversal peak (\textbf{fig. \ref{fig:current_response}b}), although the response's do differ in several ways. For \textbf{1} the tilt peak initially occurs at a more negative electric field (0.43 V/$\mu$m) than for \textbf{2} (0.1 V/$\mu$m) and is distinctly separate from the polarisation reversal peak. For \textbf{2} the peak appears the grow out of the main reversal peak seen for the N$_\text{TBF}$ phase as it moves to increasingly negative voltages. We interpret this as a consequence of their different phase transitions. Presumably for \textbf{1} in the SmC$_\text{P}^\text{H}$ phase the viscosity is similar to other smectic phases even close to the SmA$_\text{F}$ - SmC$_\text{P}^\text{H}$ transition while for \textbf{2} the viscosity close to the N$_\text{TBF}$ - SmC$_\text{P}^\text{H}$ transition is lower having just transitioned from the lower viscosity nematic phase. For both samples the peaks develop very similarly. Both the tilt peak and the polarisation reversal peak broaden and move to longer timescales, though this timescale shift is much more apparent for the polarisation reversal peak which may indicate that the tilt peak is less affected by the viscous-dynamics of the phase.

The electro-optic response was measured simultaneously to the current response by measuring the transmission of 589 nm light through the sample set between crossed polarisers (\textbf{fig. \ref{fig:current_response}}). Again, as for the current response, there are similarities and differences. Both samples show changes in the transmission that couple to the tilt peak or the polarisation reversal peak prohibiting either peak being predominately due to ion flow and accumulation. The electro-optic response is horn like for both samples although clearly much sharper for \textbf{2} indicative of the lower field thresholds for the optical changes resulting in a narrowed peak. The horn shape itself can be understood by considering that transmission through a birefringent LC slab is given by:
\begin{equation}
    \frac{T}{T_o}=\sin^2 2\phi\sin^2\left(\frac{\pi \Delta nd}{\lambda}\right)
\end{equation}
where $T/T_o$ is the transmission ratio, $\phi$ is the angle between director and polarisor, $\Delta n$ is the birefringence, $d$ the cell thickness and $\lambda$ the wavelength of light. In the SmA$_\text{F}$ state at high electric field, the transmission is a minimum as the sample has homeotropic alignment and the effective birefringence of the sample is close to zero. Reducing the electric field leads to reformation the tilt and regaining of the SmC$_\text{P}^\text{H}$ state. The out of plane field means the homeotropic alignment is retained, and so the effective birefringence is low, leading to high transmission. As the field moves towards 0 V/$\mu$m the homeotropic alignment is lost resulting in an increased birefringence and a loss in transmission. Increasing the field beyond the polarisation reversal peak results in a flip of the molecules and a field induced phase transition into the SmA$_\text{F}$ state, reducing the birefringence to zero. It does not seem that the polarisation reversal motion from a SmC$_\text{P}^\text{H}$ state to SmA$_\text{F}$ state involves an intermediate SmC$_\text{P}^\text{H}$ state with reversed polarisation.

\begin{figure}[t]
    \centering
    \includegraphics[width=1\linewidth]{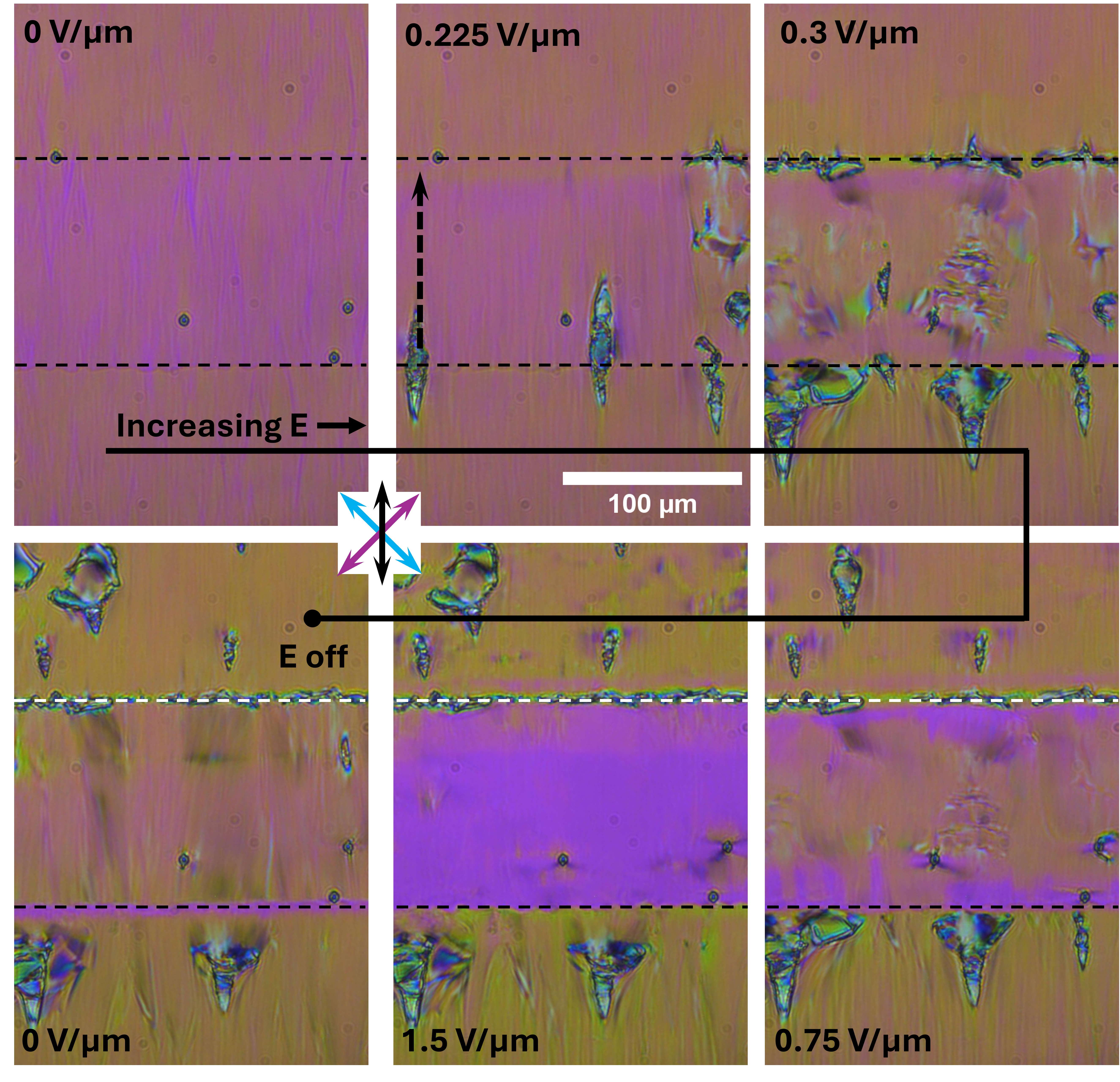}
    \caption{POM images of compound \textbf{1} at 75 °C in the SmC$_\text{P}^\text{H}$ phase in an IPS cell under applied field. The horizontal dashed lines indicate the limits of the active area while the vertical dashed arrow indicates the electric field direction. The cell is 5.5 $\mu$m thick with parallel rubbed planar alignment. The gap between the two  electrodes is 100 $\mu$m and the DC field is initially applied anti-parallel the the initial polarisation state. For all POM images polarisor axis' and rubbing direction are marked by the blue, purple and black arrows respectively.}
    \label{fig:POM_Cmp1}
\end{figure}

For the N$_\text{TBF}$ phase the current response does not show a tilt peak and only shows a single polarisation reversal peak (\textbf{fig. \ref{fig:current_response}c}). However, it has been shown that the tilt in the N$_\text{TBF}$ phase can be manipulated \cite{2024_Karcz} making the lack of an obvious tilt peak surprising. This will be revisited in section \ref{sec:ferri}. The electro-optic response is also strange with no significant change in transmission as a function of the applied triangle wave. We expect that at high electric field the molecules are in the homeotropic state giving low transmission where only very small fields are required to maintain this state. At low fields the tilt is regained along with the helix resulting in reduced transmission which counteracts the increased transmission due to increased birefringence and loss of the homeotropic state. 

\subsection{In-Plane Electro-optics of the SmC$_\text{P}^\text{H}$ Phase}

To confirm that the tilt peak is indeed associated with the reformation of molecular tilt POM observation and birefringence measurements of the in-plane switching mechanism is used. Due to the strong formation of defects in the SmC$_\text{P}^\text{H}$ of \textbf{2} (discussed in section \ref{sec:POM}) we instead will mostly focus on the SmC$_\text{P}^\text{H}$ of \textbf{1}. \textbf{1} was filled into parallel rubbed 5.5 $\mu$m thick cell with two in-plane electrodes with a gap of 100 $\mu$m. The rubbing direction was parallel to the direction of the electric field across the electrode gap.

Applying a weak electric field (0.225 V/$\mu$m) opposite to the initial director orientation results in the formation of elongated diamond shaped domains of reversed director (\textbf{fig. \ref{fig:POM_Cmp1}}). These domains grow perpendicular to the applied field until the entire active area is filled (0.3 V/$\mu$m). They can extend beyond the active area similar to the N$_\text{F}$ phase and in direct contrast to that of the SmA$_\text{F}$ phase where polarisation reversal is confined to the active area \cite{2022_Chen2} (\textbf{fig. \ref{fig:POM_Cmp1}}). This phenomena is likely due to the reduced field threshold for polarisation reversal of the SmC$_\text{P}^\text{H}$ phase vs the SmA$_\text{F}$ phase presumably allowing the resulting depolarisation field to propagate beyond the active area. We note that the diamond domains of reversed polarisation that form outside the active area tend to form on the side opposite to the direction of the applied field, though this does not mean that cannot form in the region in the direction of applied field. Increasing the field beyond the switching threshold sees a continuous gain in birefringence and a removal of the striations discussed in \ref{sec:POM} (\textbf{fig. \ref{fig:POM_Cmp1}}). The final texture is uniform in colour with a birefringence value similar to that of the SmA$_\text{F}$ phase at the SmA$_\text{F}$-SmC$_\text{P}^\text{H}$ transition. Flickering undulations in the texture can be observed indicating some hydrodynamic flow due to the applied field but these do not have any obvious pattern to them. The edges of the active area show slightly different birefringence colours due to the non-uniformity of the field in those areas. Removal of the field results in a texture that is reminiscent of the initial texture though some of the reversed polarisation domains remain outside of the active area as well as a domain line along the edges of the active area indicating the different polarisation domains.

\subsection{In-Plane Electro-optics of the N$_\text{TBF}$ Phase}

The current response of the SmC$_\text{P}^\text{H}$ phase clearly shows a second peak which we have shown to be related to tilt reformation indicating that the larger polarisation reversal peak also contains a field induced phase transition back into the SmA$_\text{F}$ phase. The N$_\text{TBF}$ phase is also heliconical but its current response does not show a second peak to associate with any field induced transition. To investigate whether full tilt removal is still possible we probe the N$_\text{TBF}$ phase via in-plane switching observations in the same 5.5 $\mu$m thick parallel rubbed cells as for the SmC$_\text{P}^\text{H}$ phase.

\begin{figure}[t]
    \centering
    \includegraphics[width=1\linewidth]{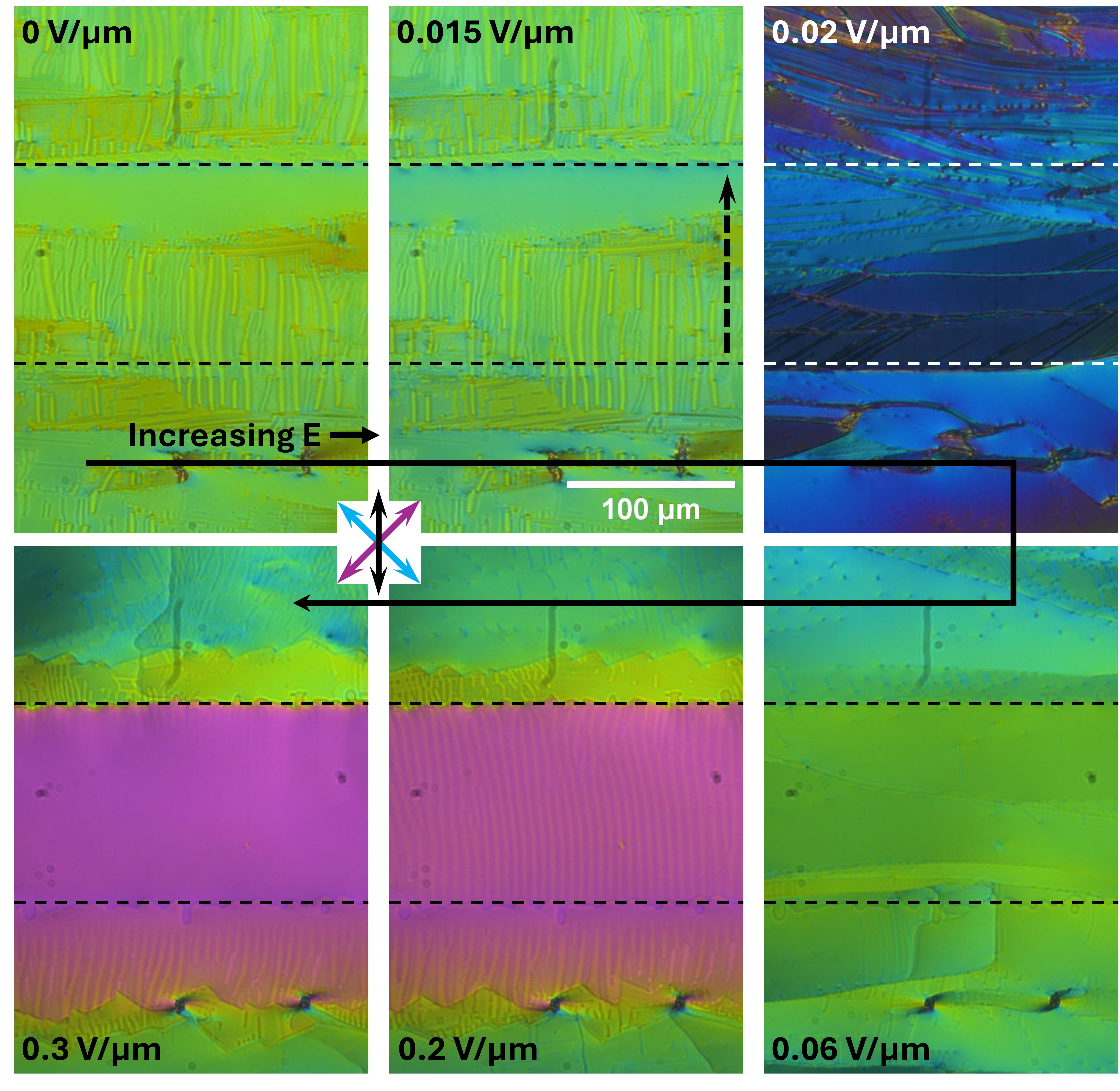}
    \caption{POM images of compound \textbf{2} at 100 °C in the N$_\text{TBF}$ phase in an IPS cell under application of a DC electric field. The horizontal dashed lines indicate the limits of the active area while the vertical dashed arrow indicates the electric field direction. The cell is 5.5 $\mu$m thick with parallel rubbed planar alignment. The gap between the two electrodes is 100 $\mu$m and the DC field is initially applied anti-parallel the the initial polarisation state. For all POM images polarisor axis' and rubbing direction are marked by the blue, purple and black arrows respectively.}
    \label{fig:POM_Cmp2}
\end{figure}

The transition to the from the N$_\text{F}$ phase to N$_\text{TBF}$ phase is clear by the formation of stripes parallel to the rubbing direction (\textbf{fig. \ref{fig:POM_Cmp2}}). As the temperature is decreased these stripes split along the rubbing direction into regions of stripe domains and uniform texture. Since these stripes are parallel to the rubbing direction and thus parallel to the helicoidal axis, they are not pitch repeat units. These are not regions of opposite polarity as they remain under application of a moderate in-plane dc field and a strong field removes them without evidence of polarisation reversal.

\begin{figure*}[t]
    \centering
    \includegraphics[width=1\linewidth]{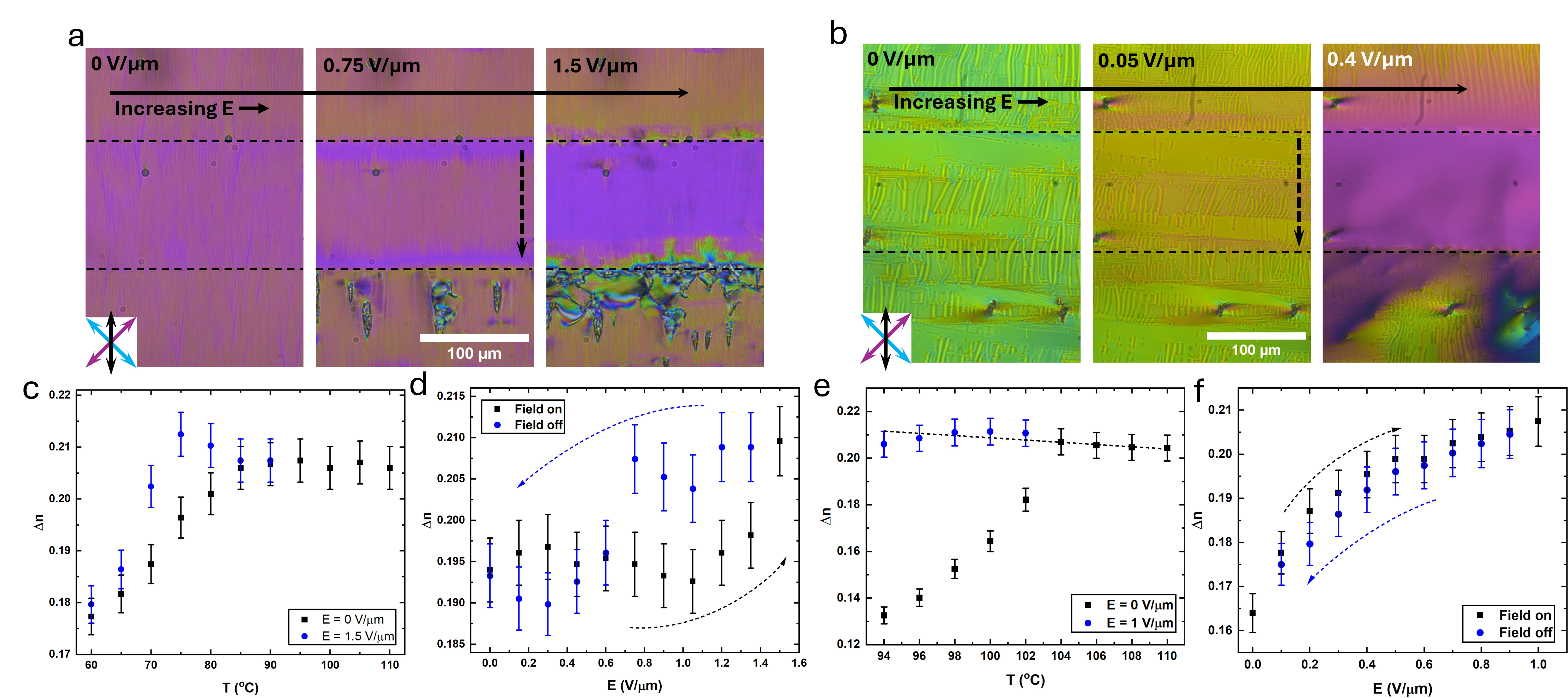}
    \caption{POM images of \textbf{a)} compound \textbf{1} at 75 °C in the SmC$_\text{P}^\text{H}$ phase and \textbf{b)} compound \textbf{2} at 100 °C in the N$_\text{TBF}$ phase both an IPS cell under application of a DC electric field where the cell is 5.5 $\mu$m thick with parallel rubbed planar alignment. Here the DC field is applied syn-parallel to the polarisation direction. For all POM images polarisor axis' and rubbing direction are marked by the blue, purple and black arrows respectively. Temperature dependence of the birefringence of \textbf{c)} \textbf{1} and \textbf{e)} \textbf{2} both with and without an applied DC field. Electric field dependence of the birefringence of \textbf{d)} compound \textbf{1} at 75 °C and \textbf{f)} compound \textbf{2} at 100 °C indicating the hysteresis of the tilt removal for the SmC$_\text{P}^\text{H}$ phase and the lack of hysteresis for the N$_\text{TBF}$ phase.}
    \label{fig:Field_Biref}
\end{figure*}

For the N$_\text{TBF}$ phase the direction of the in-plane polarity can be switched by a dc in-plane electric field just as for the N$_\text{F}$ \cite{2020_Chen, 2021_Chen, 2021_Sebastian, 2022_Basnet} although the switching is somewhat different. At low dc fields (below $\sim$ 0.02 V/$\mu$m) there is a small reduction in birefringence without significant disruption of the texture (\textbf{fig. \ref{fig:POM_Cmp2}}), consistent with the N$_\text{F}$ phase where this behaviour has been associated with increased director twist \cite{2022_Basnet}. At a critical voltage ($\sim$ 0.02 V/$\mu$m) there is a significant reduction in birefringence and the sample breaks into many domains (\textbf{fig. \ref{fig:POM_Cmp2}}). Slight uncrossing of the polarisers does not reveal any difference in optical activity of these domains. As the voltage is increased further (up to 0.06 V/$\mu$m) these domains coalesce and there is a steady increase in birefringence up to the amount of the unperturbed N$_\text{TBF}$ state with no signs of the undulations found in that unperturbed state (\textbf{fig. \ref{fig:POM_Cmp2}}). Further increase in the voltage sees an increase in the birefringence beyond the unperturbed state and eventually (above 0.15 V/$\mu$m) the undulations return before finally (around 0.3 V/$\mu$m) obtaining a uniform N$_\text{F}$ like texture with a birefringence similar to that of the N$_\text{F}$ phase just before transition to the N$_\text{TBF}$ phase. Just as for the SmC$_\text{P}^\text{H}$ phase the effect of the in-plane field can extend beyond the active area. However, due to the lower threshold for switching in the N$_\text{TBF}$ phase, the full polarisation reversal spreads significantly beyond the active area and it is just the increase of birefringence that is limited to the out of active area region opposite the applied field. It is clear that both the SmC$_\text{P}^\text{H}$ and N$_\text{TBF}$ phases can undergo a field induced phase transition the untitled state which by necessity would remove the helix entirely.

\subsection{Ferrielectric Switching in the SmC$_\text{P}^\text{H}$ Phase} \label{sec:ferri}

\begin{figure*}
    \centering
    \includegraphics[width=0.85\linewidth]{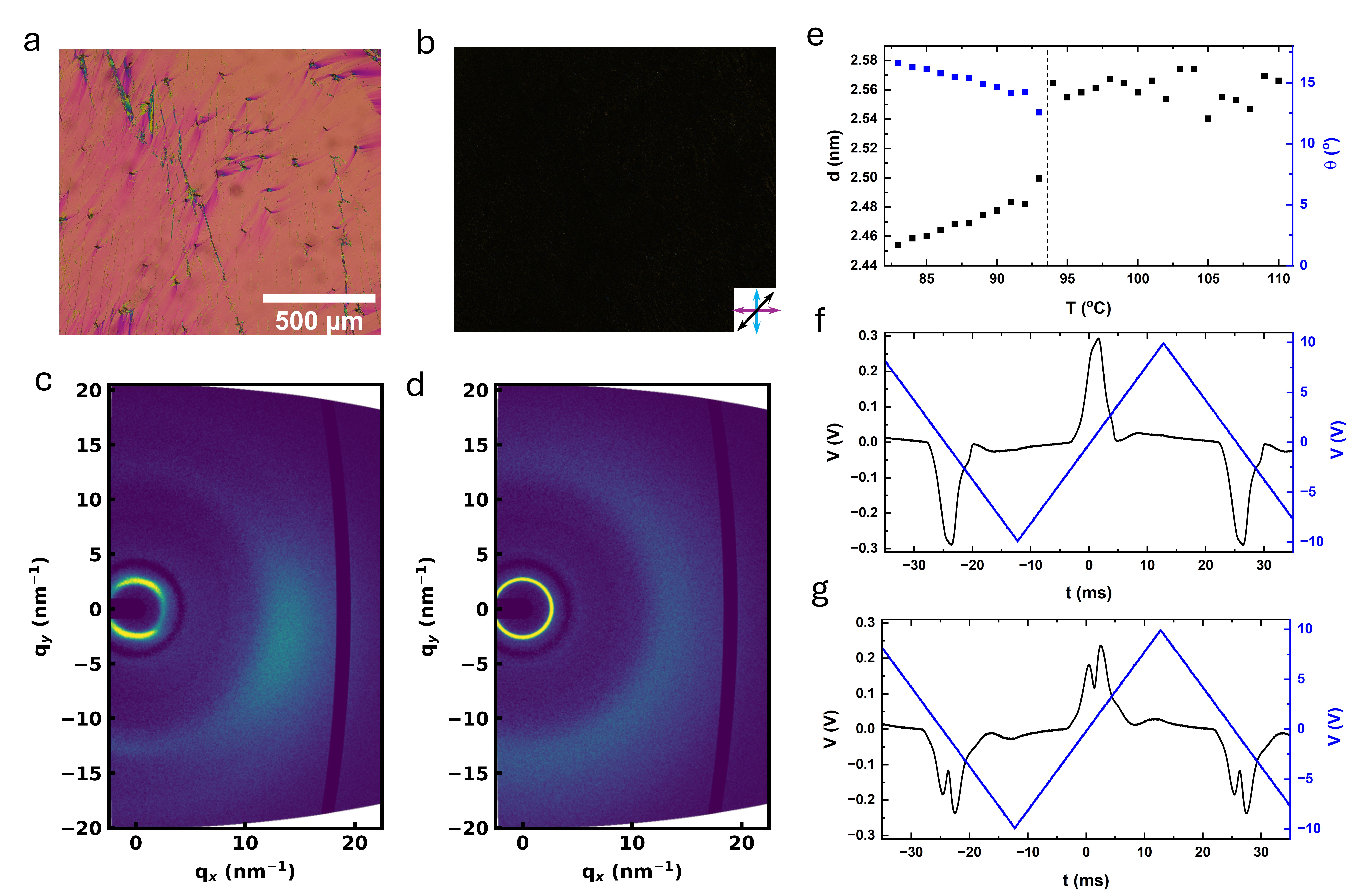}
    \caption{POM images in the \textbf{a)} N$_\text{F}$ and \textbf{b)} SmC$_\text{P}^\text{H}$ phases of \textbf{3} in a 5 $\mu$m planar aligned parallel rubbed cell. Arrows indicate direction of polarisers and rubbing direction. 2D X-ray scattering images of \textbf{3} in the \textbf{c)} N$_\text{F}$ and \textbf{d)} SmC$_\text{P}^\text{H}$ phases. \textbf{e)} Temperature dependence of the d-spacing and molecular tilt of \textbf{3}. Current responses for \textbf{3} in the \textbf{f)} N$_\text{F}$ and \textbf{g)} SmC$_\text{P}^\text{H}$ phases.}
    \label{fig:cmd3_1}
\end{figure*}

Applying a electric field anti-parallel to the polarisation axis of both the SmC$_\text{P}^\text{H}$ and N$_\text{TBF}$ phases results in full reversal of the polarisation direction. Further increase in the field results in an increase in birefringence where this second stage is due to the reduction in molecular tilt. If the electric field is applied syn-parallel to the polarisation direction no reversal in the polarisation direction is observed. However, an increase in birefringence, associated with the tilt removal, is still observed for both phases (\textbf{fig. \ref{fig:POM_Cmp2}}). If the field is removed the tilt is reformed and a drop in birefringence back to the original value is observed. For the SmC$_\text{P}^\text{H}$ phase of \textbf{1}, full tilt removal could only be completed in a specific temperature window close to the SmA$_\text{F}$ - SmC$_\text{P}^\text{H}$ transition (\textbf{fig. \ref{fig:Field_Biref}a}) due to an increase in field threshold for this effect. Presumably if higher electric fields were achieved, either through higher voltages or a narrower active area, full tilt removal could be completed for the entire phase window. For the N$_\text{TBF}$ phase of \textbf{2} full tilt removal could be completed for the entire phase (\textbf{fig. \ref{fig:Field_Biref}b}). We also note that if an in-plane field is applied to the SmC$_\text{P}^\text{H}$ phase of \textbf{2} a uniform birefringent texture could be obtained(\textbf{fig. S3}) which reverted back to the defect texture upon removal of the field indicating that tilt removal is possible here too. The hysteresis of tilt removal and switching behaviour is clear also in out of plane measurements (\textbf{fig. S4}) where tilt is clearly recovered only at lower field strength values than it is lost.

To investigate the threshold and field dependence of the tilt removal we measure birefringence as a function of applied field. This is measured in the very centre of the active area where the in-plane field is most parallel to the cell plane. For the SmC$_\text{P}^\text{H}$ phase of \textbf{1} we see that the birefringence only increases after some critical value where it sharply increases up to a value slightly beyond that of the untitled SmA$_\text{F}$ phase. We attribute this increase beyond the SmA$_\text{F}$ value either due to a increase in alignment quality over that of the SmA$_\text{F}$ or some electrically modified order parameter effect \cite{2024_Thapa}. When the field is reduced the increased birefringence is retained to lower voltages than the threshold for tilt removal indicating a hysteresis in the tilt removal (\textbf{fig. \ref{fig:Field_Biref}c}). For analogous measurements on the N$_\text{TBF}$ phase of \textbf{2} the tilt removal is threshold-less, despite the presence of an alignment layer, and has no evidence of hysteresis (\textbf{fig. \ref{fig:Field_Biref}d}). The threshold-less nature of the change in tilt is due to the avoidance of bound charges that would form in the bulk. Instead, the tilt is maintained homogenous across the active area and bound charges are expelled to the electrodes, similar to director oscillations in the N$_\text{F}$ phase \cite{2025_Basnet}. The obtained birefringence is slightly higher than the N$_\text{F}$ phase however not as significant as for the SmC$_\text{P}^\text{H}$ phase. 

The helix in the the SmC$_\text{P}^\text{H}$ and N$_\text{TBF}$ phases allows for partial polarisation compensation and so removal of the tilt results in an increase in polarisation. A ferroelectric phase has two stable states at $E=0$ while an anti-ferroelectric phase would only have a single stable state \cite{2014_Lagerwall}. A ferrielectric phase has two stable states at $E=0$ but two separate sub-lattices of non-equal polarisation \cite {2020_Fu}. These sub-lattices can be reversed going through a field-induced phase transition to the ferroelectric state at high electric field. As a result a ferrielectric phase has three hysteresis loops in its P-E curve

The SmC$_\text{P}^\text{H}$ phase has three separate hysteresis components (polarisation reversal and then tilt hysteresis in both the positive and negative polarisation states) and so has a pseudo-ferrielectric switching response. However, it is not a true ferrielectric phase as its ground state is ferroelectric like with all of its constituent dipoles orientated along the same direction. The N$_\text{TBF}$ phase has the same ground state as the SmC$_\text{P}^\text{H}$ but does not exhibit hysteresis in its tilt removal and so only exhibits ferroelectric switching with an additional pseudo-paraelectric component at higher electric fields.

Beyond the N$_\text{TBF}$ phase, current responses consisting of a small peak pre-polarity reversal and a subsequent larger peak have been observed for the non-helical SmC$_\text{P}$ phase \cite{2024_Hobbs2, 2024_Strachan, 2024_Weglowska}, though this is apparently not always observed \cite{2024_Kikuchi}. This peak has been associated with tilt reformation and indicates the possibility of a pseudo-ferrielectric responses for the the non-helical SmC$_\text{P}$ phase.

\section{Pre-Transitional Helix Formation} \label{sec:2nd_materials}

We have demonstrated how variation of the phase ordering from SmC$_\text{P}^\text{H}$-N$_\text{TBF}$-N$_\text{F}$ to SmC$_\text{P}^\text{H}$-SmA$_\text{F}$ varies the properties of the SmC$_\text{P}^\text{H}$ phase. Both of the aforementioned transitions to heliconic order are 2nd order indicated by their lack of enthalpy from DSC and their continuous, rather than discontinuous, tilt dependence at the transition to heliconic order (\textbf{fig. S11}). Variation in fluorination is a now well established technique to obtained various polar phases \cite{2025_Cruickshank, 2025_Gibb, 2025_Strachan} and so changing the chemical structure to that of compound \textbf{3} (\textbf{tbl. \ref{tbl:structures}}) results in a phase order of N$_\text{F}$-SmC$_\text{P}^\text{H}$. POM observations reveal the SmC$_\text{P}^\text{H}$-N$_\text{F}$ transition on cooling occurs as a wave edged with dendritic tendrils (\textbf{fig. \ref{fig:cmd3_1}a,b}). The transition front moves across the cell, forming a similar scattering texture as observed for \textbf{2}, nucleating from specific points across the cell. This is contrary to the smooth uniform change in texture as observed for the N$_\text{TBF}$-N$_\text{F}$ transition of \textbf{2}. On heating the scattering texture melts with a sharp boundary between SmC$_\text{P}^\text{H}$ and N$_\text{F}$ domains.

X-ray scattering measurements show a large jump in tilt angle formation up to $\sim$ 12 ° molecular tilt immediately at the transition indicating a strong first order character (\textbf{fig. \ref{fig:cmd3_1}c-e}). We also observe a comparatively large enthalpy compared to the lack of enthalpy for the 2nd order transitions SmC$_\text{P}^\text{H}$-SmA$_\text{F}$ and N$_\text{TBF}$-N$_\text{F}$ as well as the 1st order SmC$_\text{P}^\text{H}$-N$_\text{TBF}$ transition. All of this points to the SmC$_\text{P}^\text{H}$-N$_\text{F}$ transition of \textbf{3} as being the first 1st order transition to spontaneous heliconic order. Current response measurements show a single polarisation reversal peak for the N$_\text{F}$ phase (\textbf{fig. \ref{fig:cmd3_1}f}). Upon transition into the SmC$_\text{P}^\text{H}$ phase the peak splits into two with a smaller tilt peak at pre-polarity voltages and larger polarisation reversal peak (\textbf{fig. \ref{fig:cmd3_1}g}) which moves to longer time scales with reducing temperature. The larger initial size of the tilt peak is owed to the first order nature of the transition leading to an discontinuous large change in the molecular tilt and so a larger change in polarisation upon tilt reformation. 

\begin{figure}
    \centering
    \includegraphics[width=1\linewidth]{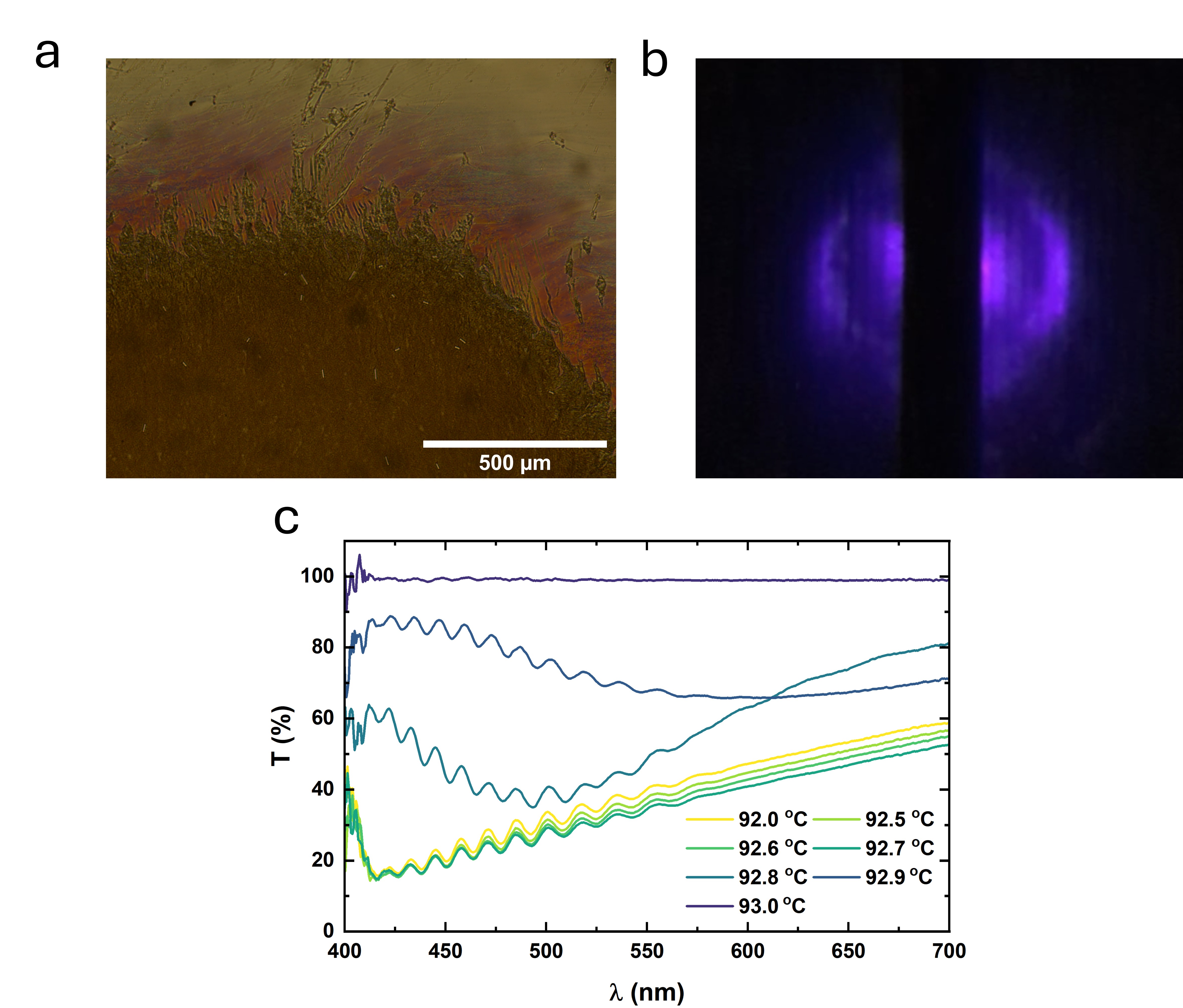}
    \caption{\textbf{a)} Microscopy image with the polarisor inserted parallel to the rubbing direction (no analyser) of the boundary between the N$_\text{F}$ phase (top) and SmC$_\text{P}^\text{H}$ phase (bottom) showing selective reflection in the boundary region. \textbf{b)} diffraction pattern from the boundary region demonstrating a pitch of $\sim$ 800 $\mu$m. \textbf{c)} Transmission spectrum from the boundary region of image \textbf{a)}.}
    \label{fig:cmd3_2}
\end{figure}

If the boundary between the N$_\text{F}$ and SmC$_\text{P}^\text{H}$ domains is observed on cooling, a narrow region of $\sim$ 300 $\mu$m width exists where selective reflection can be seen (\textbf{fig. \ref{fig:cmd3_2}a}). Across this boundary layer the colour varies from red to purple indicating an increasing in bandgap wavelength from the SmC$_\text{P}^\text{H}$ domain to the N$_\text{F}$ domain. Measuring the wavelength of the selective reflection directly shows it does indeed shift rapidly though the visible window in 0.2 °C indicating rapid change of the helical pitch in this region (\textbf{fig. \ref{fig:cmd3_2}c}). Direct remeasurements of the pitch via laser diffraction confirm this indicating a change in pitch from 920 $\mu$m to less than 600 $\mu$m within that same 0.2 °C window (\textbf{fig. \ref{fig:cmd3_2}b}). We considered whether this is a narrow temperature window of N$_\text{TBF}$ phase existence, however on heating the region of selective reflection and light diffraction due to the helical pitch is not observed (videos 1 and 2). We interpret this as either the SmC$_\text{P}^\text{H}$ phase front growing across the cell and forming initially in a thin area nearer the base due to any vertical temperature gradients which then produces the selective reflection and pitch. Alternatively this is a pre-transitional effect where the helix in the SmC$_\text{P}^\text{H}$ acts in induce some helical ordering in the neighbouring N$_\text{F}$ regions. As this is not the lowest free energy state for the N$_\text{F}$ phase this helix unwinds as distance from the phase front is increased until it smoothly merges with the non-helical N$_\text{F}$ regions.

\section{Conclusion}

The discovery of spontaneous chiral symmetry breaking in the ferroelectric nematic realm \cite{2024_Karcz, 2024_Gibb} is extremely exciting due to the potential for replicating the behaviours seen in hard matter systems such as ferromagnets in fluid soft matter systems. Here we have demonstrated that the SmC$_\text{F}$ phase found in \textbf{2} \cite{2024_Karcz} is helical and is in fact a SmC$_\text{P}^\text{H}$ phase, as observed for \textbf{1} \cite{2024_Gibb}. We show that there is at least three phase pathways to the SmC$_\text{P}^\text{H}$ (we expect that there could well be more pathways then the three presented here). Namely SmC$_\text{P}^\text{H}$-SmA$_\text{F}$, SmC$_\text{P}^\text{H}$-N$_\text{TBF}$ and SmC$_\text{P}^\text{H}$-N$_\text{F}$ and that each leads to an array of different polymorphs and interesting properties.

We account for the double peak structure of the current response for the SmC$_\text{P}^\text{H}$ phase, demonstrating that the smaller pre-polarity peak is associated with reformation of the tilt and show how the behaviour of this tilt peak is influenced by the phase preceding it. The tilt peak of the SmC$_\text{P}^\text{H}$ phase is indicative of field induced phase transitions to the SmA$_\text{F}$ state. This phase transition shows clear hysteresis indicating that it is pseudo-ferrielectric due to the increase in polarisation that occurs by removing molecular tilt from a partially compensated helical phase.

Looking forward, spontaneous symmetry breaking due to polar order may yield interesting new phase types when superimposed on more complex liquid crystal organisations, for example the twist grain boundary like polar twisted smectic structures described recently by Nishikawa et. al. \cite{2025_Nishikawa}

\section*{Data availability}

The data associated with this paper are openly available from the University of Leeds Data Repository at: https://doi.org/10.5518/1643

\section*{Author Contribution Statement}

JH and CJG contributed equally to this work. The manuscript was written, reviewed, and edited with contributions from all authors.   

\section*{Competing Interests}

Authors declare that they have no competing interests

\section*{Acknowledgements}

RJM thanks UKRI for funding via a Future Leaders Fellowship, grant number MR/W006391/1, and the University of Leeds for funding via a University Academic Fellowship. The SAXS/WAXS system used in this work was funded by EPSRC via grant number EP/X0348011. R.J.M. gratefully acknowledge support from Merck KGaA.  Computational work was performed on ARC3 and ARC4, part of the high-performance computing facilities at the University of Leeds. 

\bibliography{aipsamp}

\end{document}


\title{Supplementary Information:\\Ferri- and Ferro-Electric Switching in Spontaneously Chiral Polar Liquid Crystals}

\author{Jordan Hobbs$^1$}
\email{j.l.hobbs@leeds.ac.uk}
\author{Calum J. Gibb$^2$}
\author{Richard. J. Mandle$^{1,2}$}
\affiliation{$^1$School of Physics \& Astronomy, University of Leeds, UK}
\affiliation{$^2$School of Chemistry, University of Leeds, UK}

\date{\today}
\maketitle

\section*{Contents}

\begin{enumerate}
\def\labelenumi{\arabic{enumi}.}
\item
  Supplementary methods
\item
  Supplementary results
\item
  Organic synthesis
\item
  Supplementary References
\end{enumerate}

\section{Supplementary Methods}

\subsection*{1.1. Chemical Synthesis}

Chemicals were purchased from commercial suppliers (Fluorochem, Merck, ChemScene, Ambeed) and used as received. Solvents were purchased from Merck and used without further purification. Reactions were performed in standard laboratory glassware at ambient temperature and atmosphere and were monitored by TLC with an appropriate eluent and visualised with 254 nm light. Chromatographic purification was performed using a Combiflash NextGen 300+ System (Teledyne Isco) with a silica gel stationary phase and a hexane/ethyl acetate gradient as the mobile phase, with detection made in the 200-800 nm range. Chromatographed materials were filtered through 200 nm PTFE frits and then subjected to re-crystallisation from an appropriate solvent system.

\subsection*{1.2. Chemical Characterisation}

Chemical materials were characterised by NMR spectroscopy using a Bruker Avance III HDNMR spectrometer operating at 400 MHz, 100.5 MHz or 376.4 MHz (1H, 13C{1H} and 19F, respectively). High resolution mass spectrometry data (HRMS) was collected using a Bruker MaXis Impact spectrometer with a negative ESI source (VIP-HESI); the sample was introduced via direct infusion as solution in MeCN. HPLC analysis was performed using an Agilent 1290 Infinity II system fitted with a poroshell 120 ec-c18 column running a water:MeCN gradient.

\subsection*{1.3. Thermal Analysis}

Differential scanning calorimetry (DSC) measurements were performed using a TA Instruments Q2000 DSC instrument (TA Instruments, Wilmslow UK), equipped with a RCS90 Refrigerated cooling system (TA Instruments, Wilmslow UK). The instrument was calibrated against an Indium standard, and data were processed using TA Instruments Universal Analysis Software. Samples were analysed under a nitrogen atmosphere, in hermetically sealed aluminium TZero crucibles (TA Instruments, Wilmslow, UK) and subjected to three analysis cycles. In all cases, samples were subject to heating and cooling at a rate of 10 K min-1.  Phase transition temperatures were measured as onset values on cooling cycles for consistency between monotropic and enantiotropic phase transitions, while crystal melts were obtained as onset values on heating. 

\subsection*{1.3. Optical Microscopy}

Polarised light optical microscopy (POM) was performed using a Leica DM2700P polarised light microscope (Leica Microsystems (UK) Ltd., Milton Keynes, UK), equipped with 10x and 50x magnification, and a rotatable stage. A Linkam TMS 92 heating stage (Linkam Scientific Instruments Ltd., Redhill, UK) was used for temperature control, and samples were studied sandwiched between two untreated glass coverslips. Images were recorded using a Nikon D3500 Digital Camera (Nikon UK Ltd., Surbiton, UK), using DigiCamControl software.  

\subsection*{1.4. X-ray Scattering}

X-ray scattering measurements, both small angle (SAXS) and wide angle (WAXS) were recorded using an Anton Paar SAXSpoint 5.0 beamline machine.
This was equipped with a primux 100 Cu X-ray source with a 2D EIGER2 R
detector. The X-rays had a wavelength of 0.154 nm. Samples were filled
into either thin-walled quartz capillaries or held between Kapton tape.
Temperature was controlled using an Anton Paar heated sampler with a
range of -10 \textsuperscript{o}C to 120\textsuperscript{o}C and the
samples held in a chamber with an atmospheric pressure of \textless1
mbar. Samples were held at 120\textsuperscript{o}C to allow for
temperature equilibration across the sample and then slowly cooled while
stopping to record the scattering patterns.

The samples were not formally aligned and so these measurements can be
considered as ``powder'' samples. It should be noted that some
spontaneous alignment of the LCs both within the capillaries and between
the Kapton tape did occur leading to the classic ``lobe'' pattern seen
in the 2D patterns. 1D patterns were obtained by radially integrating
the 2D SAXS patterns. Peak position and FWHM was recorded and then
converted into d spacing following Bragg's law. In the tilted smectic
phase, the tilt was obtained from:
\begin{equation}
    \frac{d_{c}}{d_{A}} = \cos \theta
\end{equation}

where \(d_{c}\) is the layer spacing in the tilted smectic phase,
\(d_{A}\) is the extrapolated spacing from the non-tilted preceding phase, extrapolated to account for the weak temperature dependence of the preceding phases due to shifts in conformation and order, and \(\theta\) the structural tilt angle.

\subsection*{1.5. Measurement of Spontaneous Polarization
(P\textsubscript{S})}

Spontaneous polarisation measurements are undertaken using the current
reversal technique. Triangular waveform AC voltages are applied to the
sample cells with an Agilent 33220A signal generator (Keysight
Technologies), and the resulting current outflow is passed through a
current-to-voltage amplifier and recorded on a RIGOL DHO4204
high-resolution oscilloscope (Telonic Instruments Ltd, UK). Heating and
cooling of the samples during these measurements is achieved with an
Instec HCS402 hot stage controlled to 10 mK stability by an Instec
mK1000 temperature controller. The LC samples are held in 4µm thick
cells with no alignment layer, supplied by Instec. The measurements
consist of cooling the sample at a rate of 1 Kmin\textsuperscript{-1}
and applying a set voltage at a frequency of 20 Hz to the sample every 1
K. The voltage was set such that it would saturate the measured
P\textsubscript{S} and was determined before final data collection.

There are three contributions to the measured current trace:
accumulation of charge in the cell (I\textsubscript{c}), ion flow
(I\textsubscript{i}), and the current flow due to polarisation reversal
(I\textsubscript{p}). To obtain a P\textsubscript{S} value, we extract
the latter, which manifests as one or multiple peaks in the current
flow, and integrate as:

\begin{equation}
    P_{S} = \int_{}^{}{\frac{I_{p}}{2A}dt}
\end{equation}

where A is the active electrode area of the sample cell.

\subsection*{1.6. Birefringence Measurements}

Birefringence was measured using a Berek compensator mounted in a Leica
DM 2700 P polarised optical microscope. The birefringence was measured
in the centre of the active ITO area for the IPS cells used here.

\subsection*{1.7. Transmission Spectra}

A microscope equipped coupled to an Avantes AvaSpec-2048 XL spectrometer
which used to record transmission spectra of samples. The rubbing
direction was set parallel to the light polarization direction, but no
analyzer was inserted. A spectra from the phase above the spontaneously
chiral phases was used as the reference state of each sample

\subsection*{1.8. Pitch Measurements}

The pitch was measured by illumination of the sample from the bottom via
a 405 nm laser. The angle of diffraction was measured on a flat screen
set 1.8cm away from the sample.

\section{Supplementary Results}

\begin{figure}[H]
    \centering
    \includegraphics[width=1\linewidth]{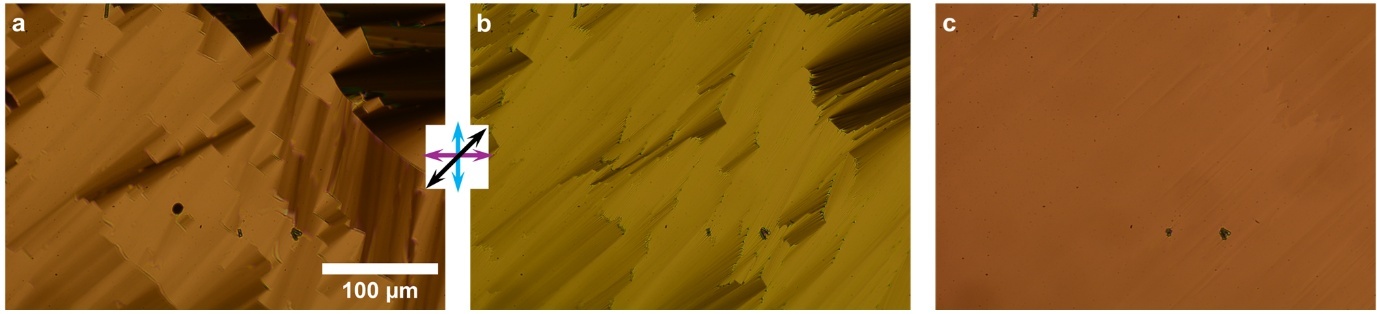}
    \caption{POM images of compound \textbf{1} in a 5 $\mu$m thick
parallel rubbed planar cell at a) 120 °C in the SmA\textsubscript{F}
phase on heating, b) 150 °C in the SmA phase on heating and c) 120 °C in
the SmA\textsubscript{F} phase on cooling. The blue and purple arrows
indicate the polariser axes while the black arrow indicates the rubbing
direction.}
\end{figure}

\begin{figure}[H]
    \centering
    \includegraphics[width=1\linewidth]{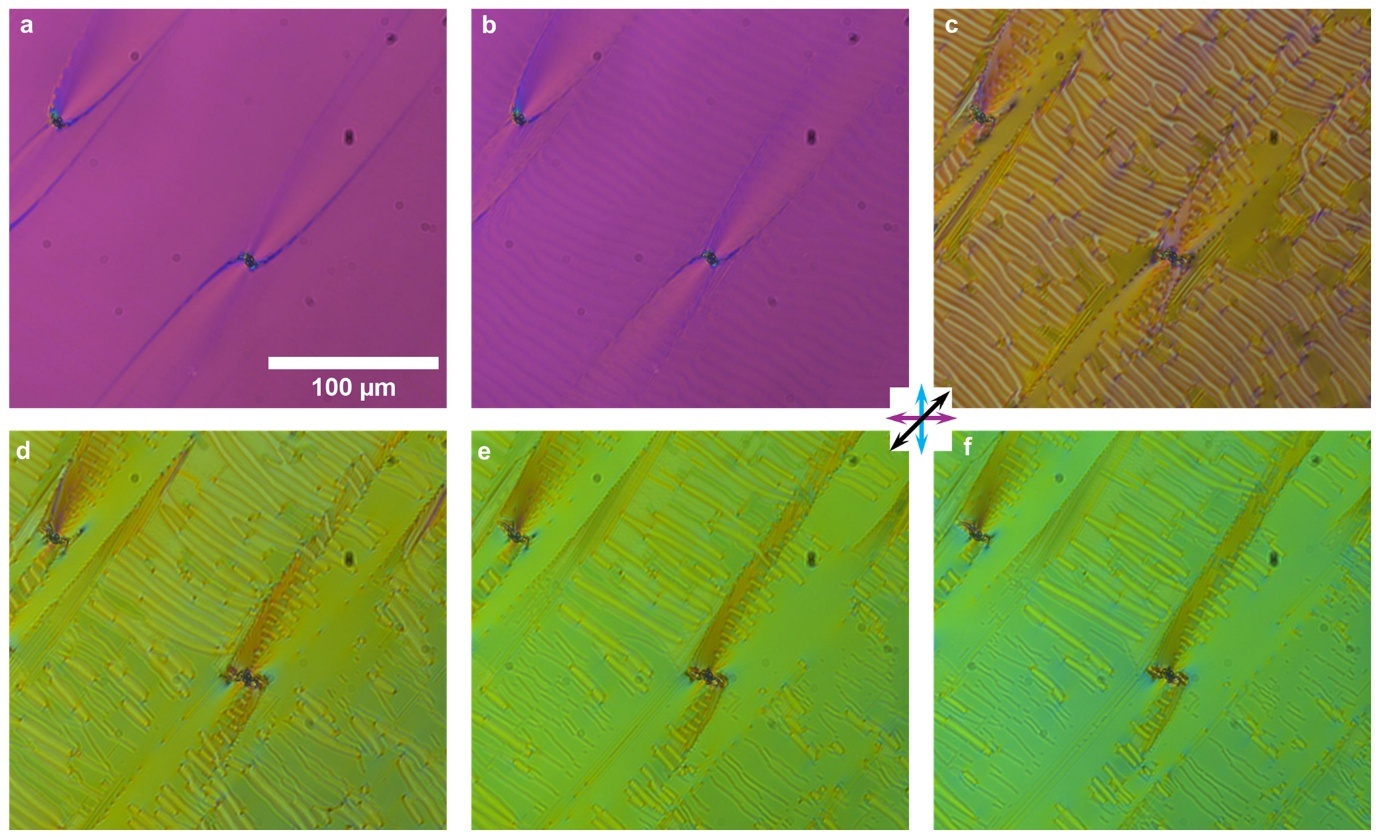}
    \caption{POM images of compound \textbf{2} in a 5 $\mu$m thick parallel rubbed
planar cell at a) 105 °C in the N\textsubscript{F} phase on cooling and
b) 104 °C, c) 103 °C, d) 102 °C e) 101 °C and f) 100 °C in the
N\textsubscript{TBF} phase on cooling. The blue and purple arrows
indicate the polariser axes while the black arrow indicates the rubbing
direction.}
\end{figure}

\begin{figure}[H]
    \centering
    \includegraphics[width=1\linewidth]{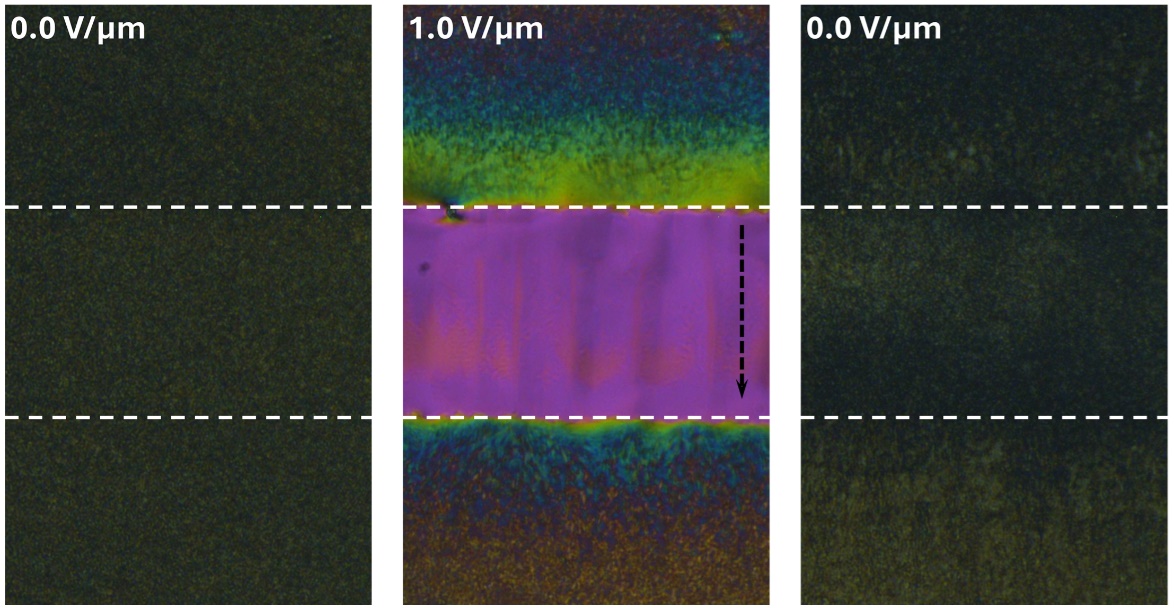}
    \caption{POM observations of compound \textbf{2} in the
\(\text{SmC}_{\text{P}}^{\text{H}}\) phase at 92 ºC in a 5.5 $\mu$m parallel
rubbed planar aligned cell with a pair of in plane electrodes separated
by 100 µm. Voltage is applied OFF-ON-OFF.}
\end{figure}

\begin{figure}[H]
    \centering
    \includegraphics[width=1\linewidth]{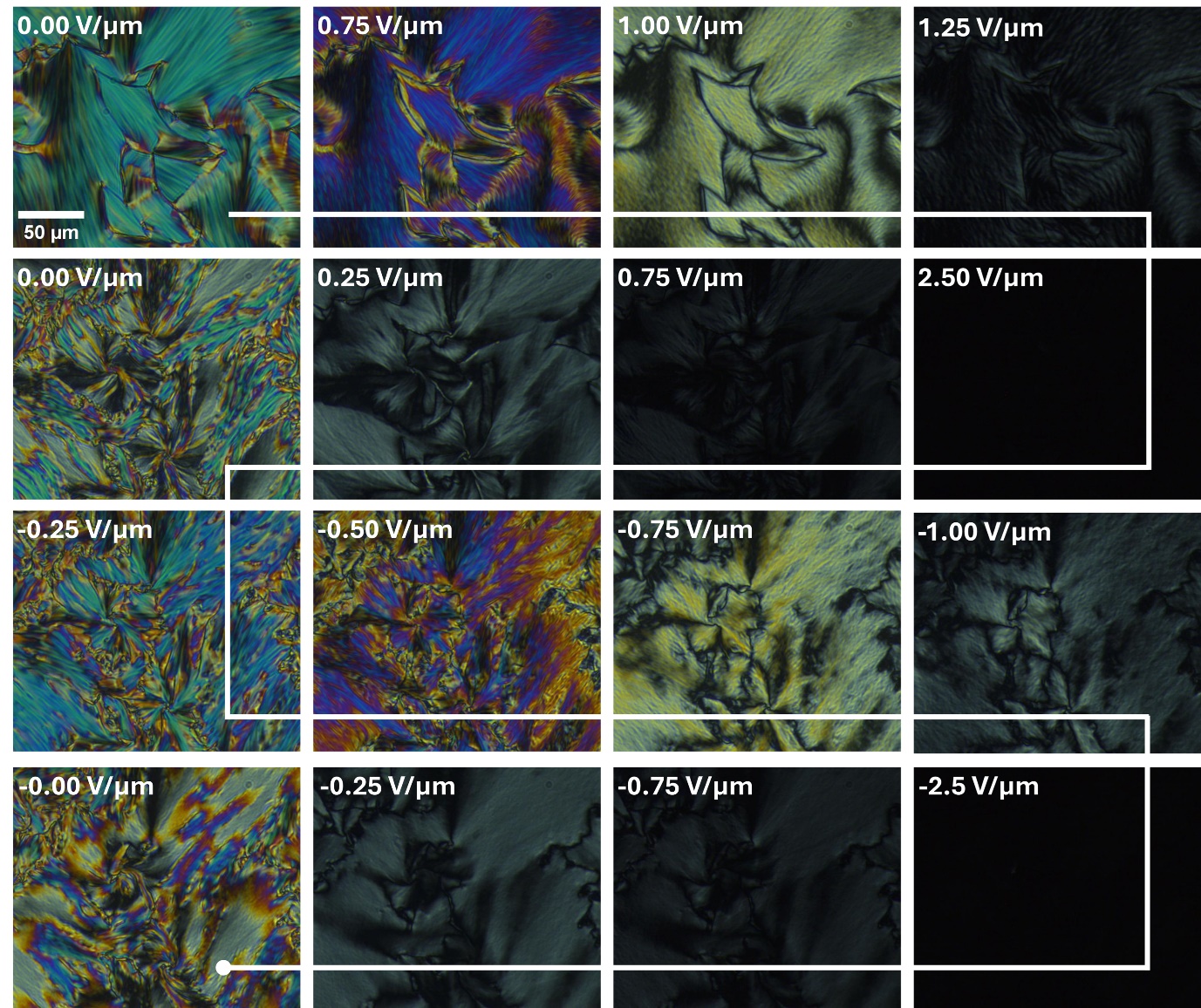}
    \caption{POM observations of compound \textbf{1} in the
\(\text{SmC}_{\text{P}}^{\text{H}}\) phase at 70 °C in a 4 $\mu$m cell with
no alignment layer. Field strength values indicate the strength of DC
field applied across the cell gap.}
\end{figure}

\begin{figure}[H]
    \centering
    \includegraphics[width=1\linewidth]{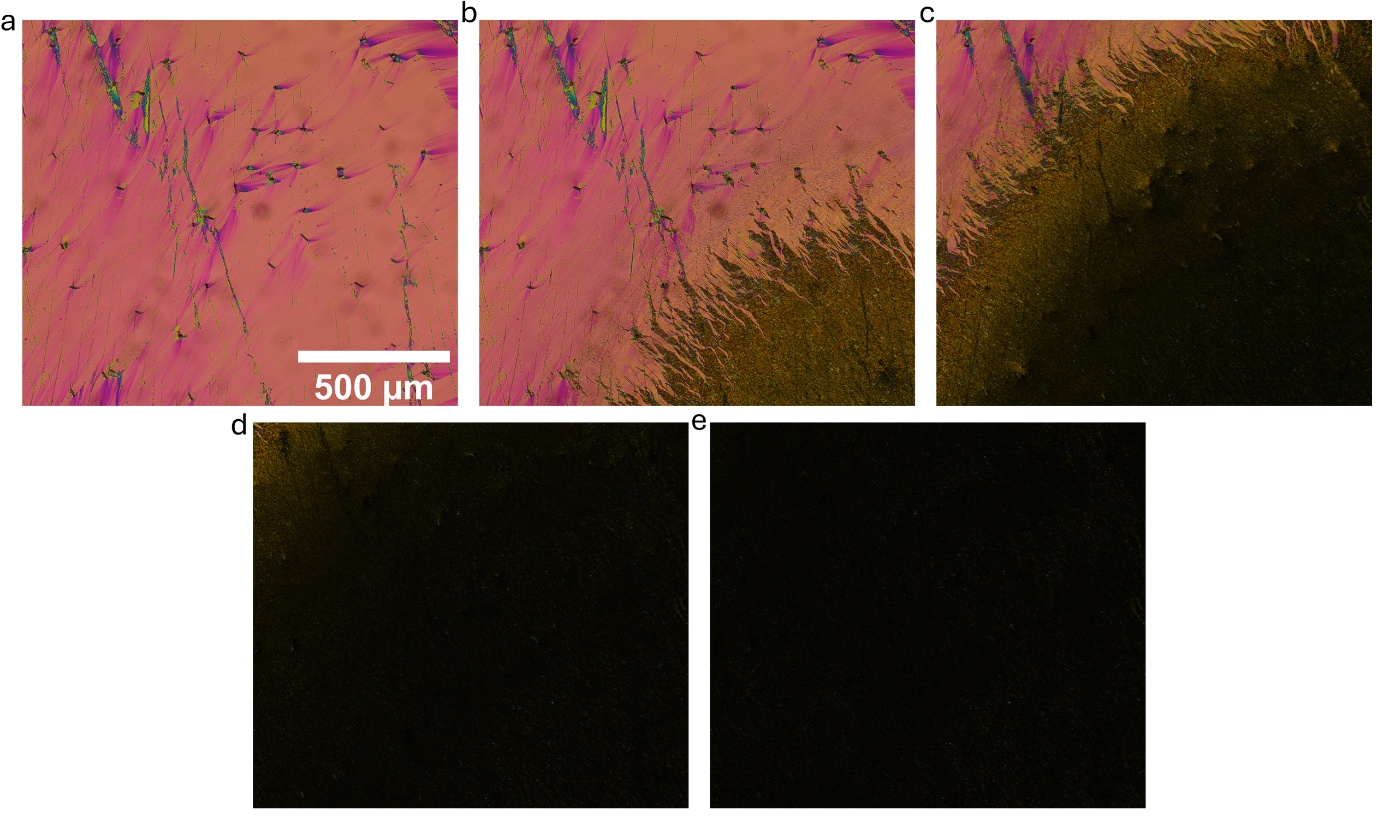}
    \caption{POM observations of \textbf{3} at the boundary between the
N\textsubscript{F} and \(\text{SmC}_{\text{P}}^{\text{H}}\) phases over
1 °C showing the slow growth and dendritic tendrils.}
\end{figure}

\begin{figure}[H]
    \centering
    \includegraphics[width=1\linewidth]{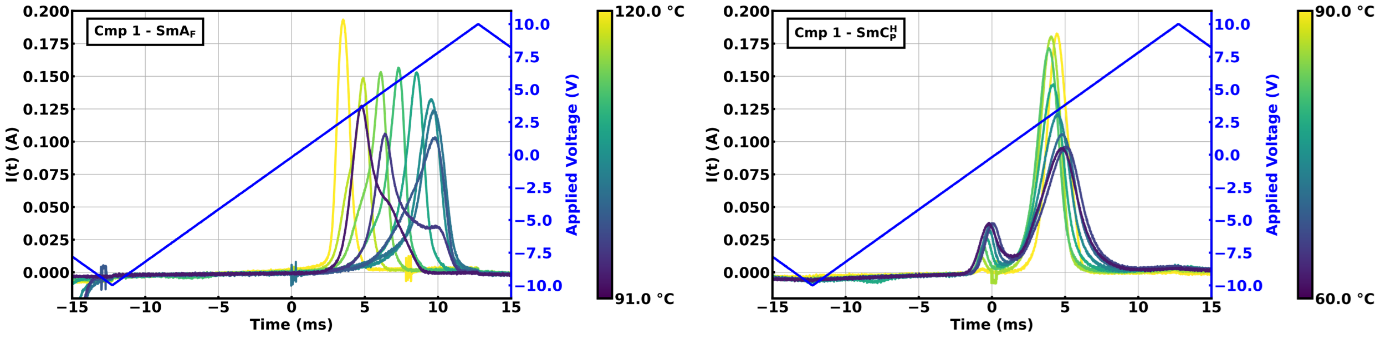}
    \caption{Current responses for compound \textbf{1} in a 4 $\mu$m cell with no alignment layer. The applied field is a triangle wave with 10 V (2.5 V/$\mu$m) at 10 Hz.}
\end{figure}

\begin{figure}[H]
    \centering
    \includegraphics[width=1\linewidth]{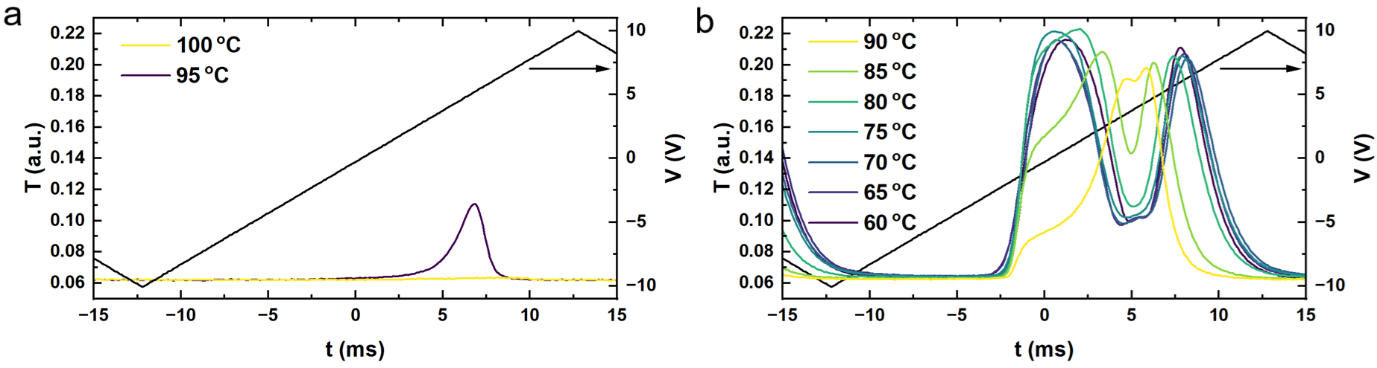}
    \caption{Electro-optic response for compound \textbf{1} in a 4 $\mu$m cell with no alignment layer set between crossed polarisers. The applied field is a triangle wave with 10 V (2.5 V/$\mu$m) at 10 Hz.}
\end{figure}

\begin{figure}[H]
    \centering
    \includegraphics[width=1\linewidth]{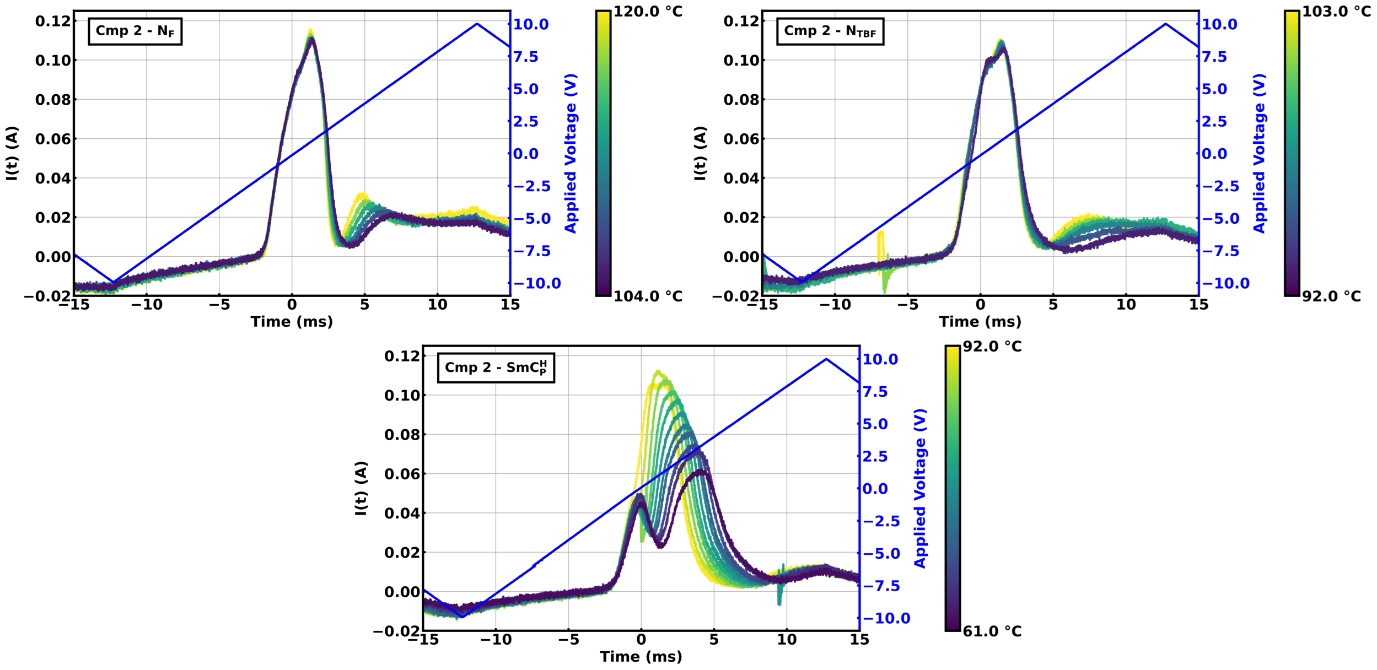}
    \caption{Current responses for compound \textbf{2} in a 4 $\mu$m cell with no alignment layer. The applied field is a triangle wave with 10 V (2.5 V/$\mu$m) at 10 Hz}
\end{figure}

\begin{figure}[H]
    \centering
    \includegraphics[width=1\linewidth]{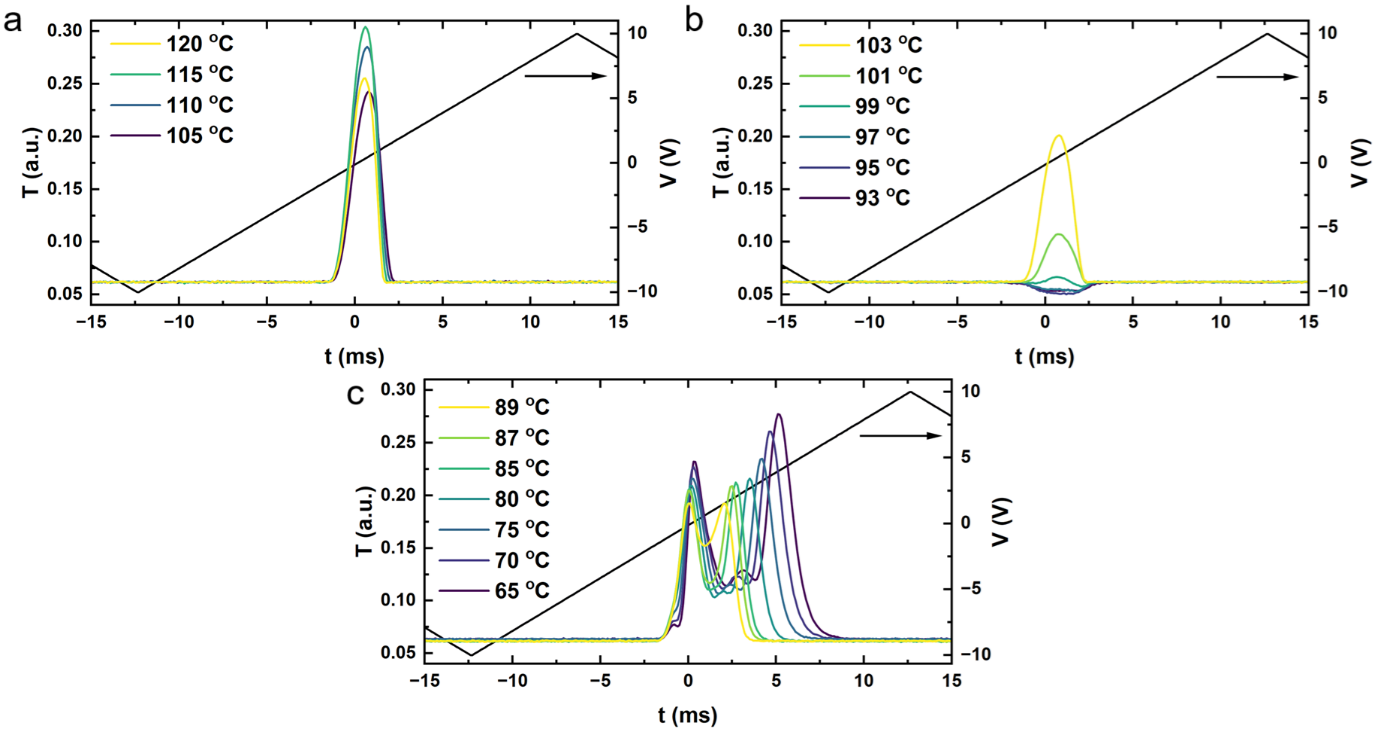}
    \caption{Electro-optic response for compound \textbf{2} in a 4 $\mu$m cell with no alignment layer set between crossed polarisers. The applied field is a triangle wave with 10 V (2.5 V/$\mu$m) at 10 Hz.}
\end{figure}

\begin{figure}[H]
    \centering
    \includegraphics[width=1\linewidth]{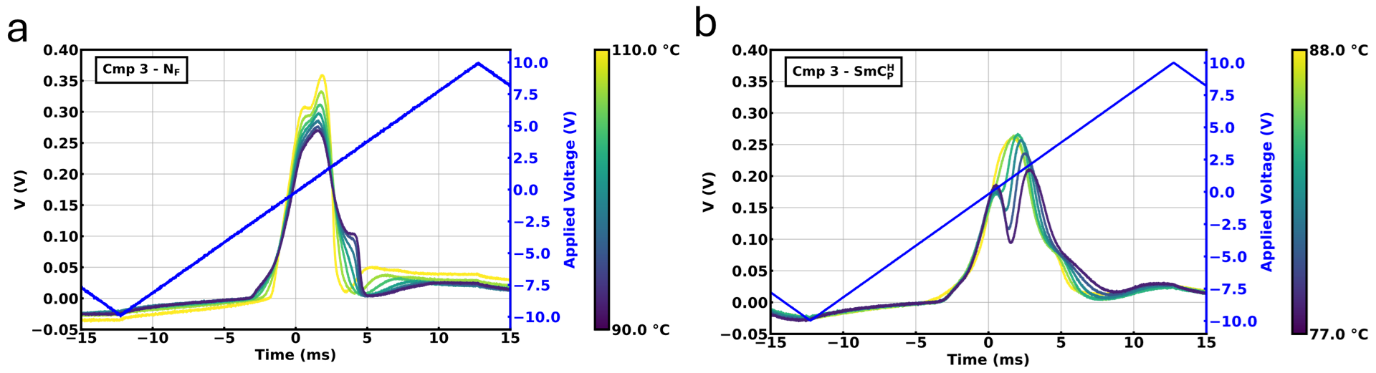}
    \caption{Current responses for compound \textbf{3} in a 4 $\mu$m
cell with no alignment layer. The applied field is a triangle wave with
10 V (2.5 V/$\mu$m) at 10 Hz}
\end{figure}

\begin{figure}[H]
    \centering
    \includegraphics[width=1\linewidth]{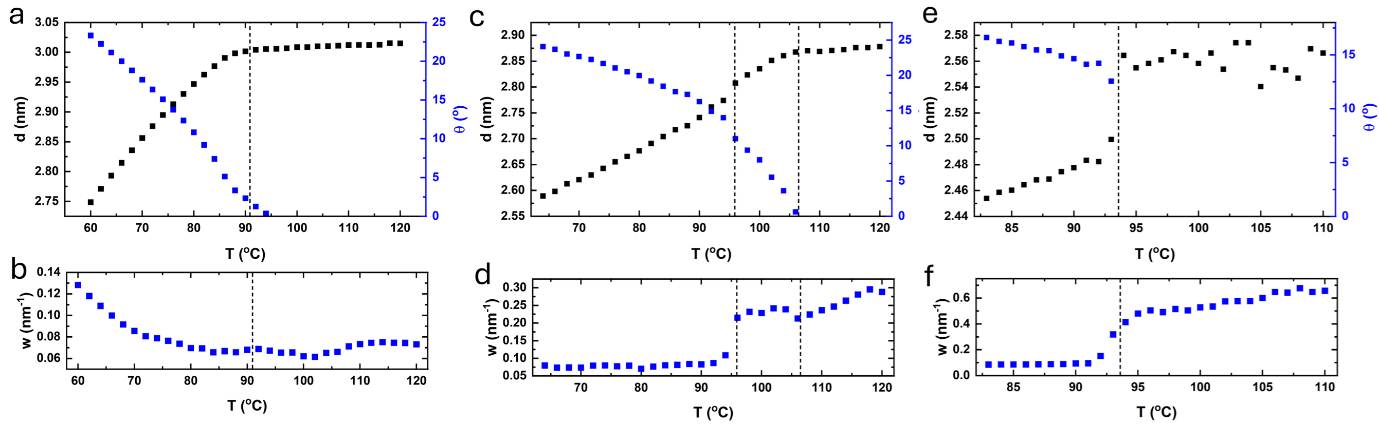}
    \caption{Temperature dependence and FWHM of the small angle scattering peaks of a,b) \textbf{1}, c,d) \textbf{2} and e,f) \textbf{3.}}
\end{figure}

\begin{figure}[H]
    \centering
    \includegraphics[width=1\linewidth]{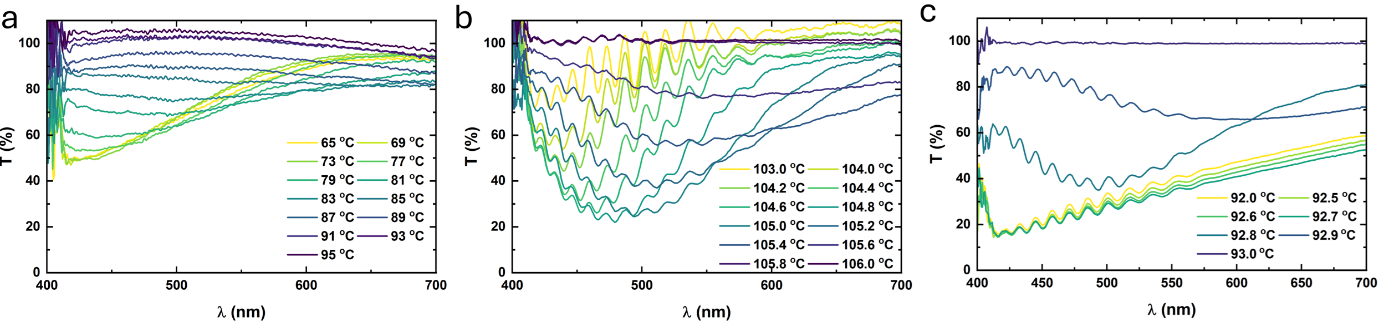}
    \caption{Transmission spectrum of a) \textbf{1} b) \textbf{2} and c) \textbf{3}. For all samples the wavelength shifts to shorter wavelengths.}
\end{figure}

\begin{figure}[H]
    \centering
    \includegraphics[width=0.5\linewidth]{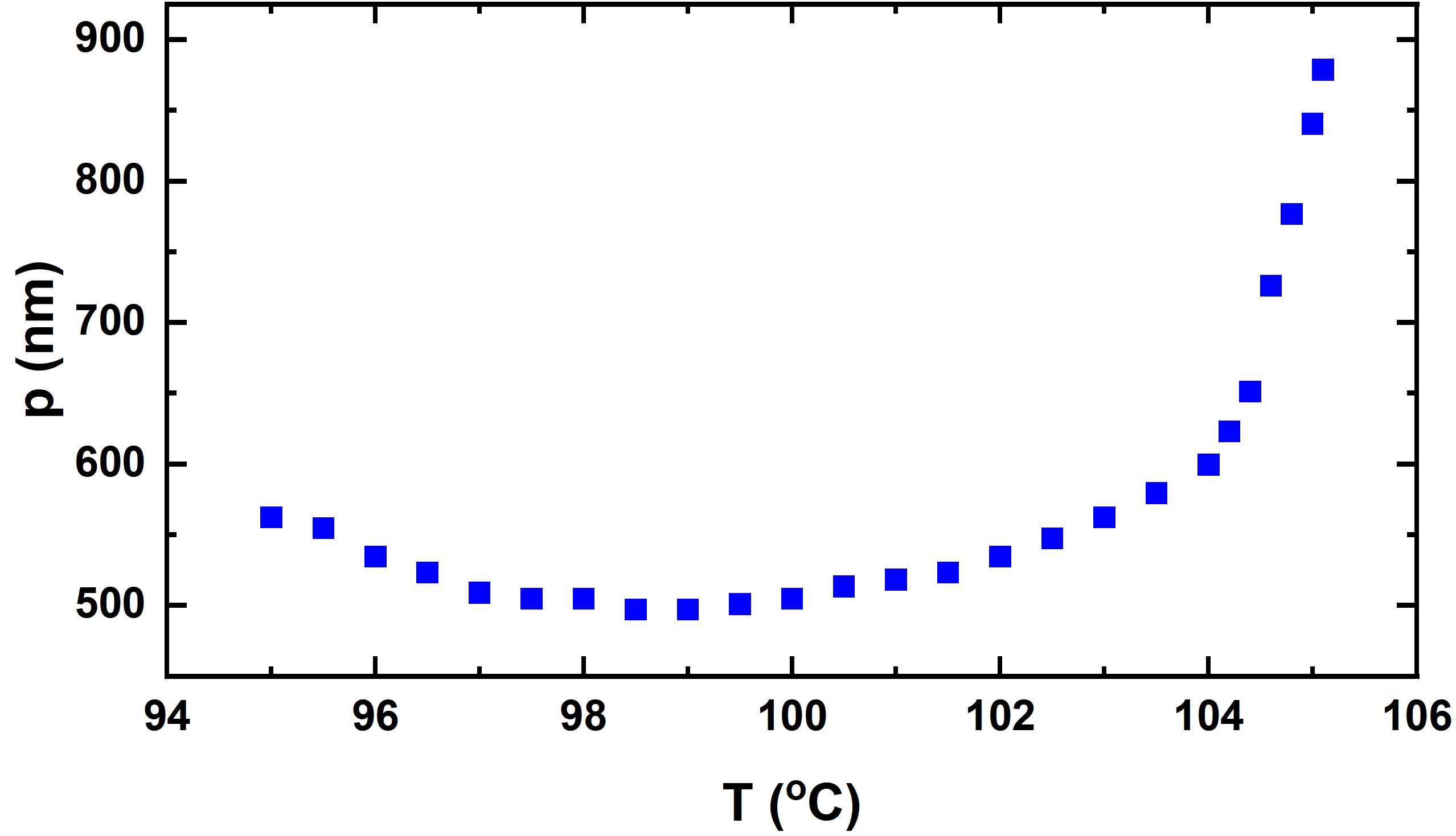}
    \caption{Temperature dependence of the pitch of \textbf{2} in the N\textsubscript{TBF} phase obtained on cooling in a 5 $\mu$m parallel rubbed planar cell.}
\end{figure}

\section{Organic Synthesis}

The synthetic route and chemical characterisation data associated with
intermediate products and materials \textbf{2-4} is given below. The
synthesis of material \textbf{1} and the preparation of
2,6-difluoro-4-(5-propyl-1,3-dioxan-2-yl)benzoic acid is reported
elsewhere {[}1,2{]}

\subsection*{3.1. General Suzuki Coupling Protocol}

A reaction flask was charged with
5-bromo-2-(difluoro(3,4,5-trifluorophenoxy)methyl)-1,3-difluorobenzene
(20 mmol) and the relevant phenol boronic ester (21 mmol) which were
dissolved in 150 mL of THF and 60 mL of 2M
K\textsubscript{2}CO\textsubscript{3}(aq). The resultant solution was
sparged with N\textsubscript{2(g)} for 20 mins. In a separate vial, 5 mL
of THF was sparged with N\textsubscript{2(g)} for 15 minutes before
Pd(OAc)2 (50 mg) and SPhos (100 mg) were added and stirred for a further
5 mins. The reaction flask was then heated to 70 ℃ and the catalyst
solution added in one portion. The reaction was monitored by TLC with
the completion of the reaction being determined by the complete
consumption of the bromo-sub-straight. The reaction was then cooled, the
aqueous and organic layers separated with the organics being dried over
MgSO\textsubscript{4}. The organics were then passed through a silica
plug before the filtrate was concentrated under reduced pressure and
purified by flash chromatography over silica gel with a gradient of
hexane/ethyl acetate using a Combiflash NextGen300+ system using a
gradient elution from hexane - ethyl acetate. Finally, the product
re-crystallised from hexane:toluene (10:1) to give the appropriate
phenols.

Insert scheme here

\textbf{Scheme 1.} General Suzuki coupling protocol.

\begin{figure}[H]
    \centering
    \includegraphics[width=0.3\linewidth]{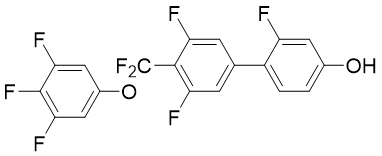}
\end{figure}

\sloppy\textbf{4-Hydroxy-5-(2-fluoro-4-hydroxyphenyl)-2-(difluoro(3,4,5-trifluorophenoxy)methyl)-1,3-difluorobenzene}

Appearance: White powder

R\textsubscript{f} (DCM): 0.35

\textsuperscript{1}H NMR (400 MHz, DMSO) ($\delta$): 10.38 (s, 1H,
Ar-O\textbf{H}), 7.50 (t, J = 8.5 Hz, 1H, Ar-\textbf{H}), 7.47 -- 7.42
(m\textsubscript{apparent}, 2H, Ar-\textbf{H}), 7.42 -- 7.32
(m\textsubscript{apparent}, 2H, Ar-\textbf{H}), 6.77 -- 6.67
(m\textsubscript{apparent}, 2H, Ar-\textbf{H}).

\textsuperscript{13}C\{\textsuperscript{1}H\} NMR (101 MHz, DMSO) ($\delta$):
160.43 (d, \emph{J} = 249.3 Hz), 159.97 (d, \emph{J} = 248.2 Hz), 157.76
(d, \emph{J} = 6.2 Hz), 150.26 (ddd, \emph{J} = 248.7, 10.5, 5.1 Hz),
144.17 (t, \emph{J} = 16.3 Hz), 142.08 (t, \emph{J} = 11.4 Hz), 137.83
(d, \emph{J} = 248.3 Hz), 131.32, 122.63, 120.00, 117.38, 115.07 (d,
\emph{J} = 11.9 Hz), 112.71, 112.49, 108.10 -- 107.67
(m\textsubscript{apparent}), 106.45 (t, \emph{J} = 13.3 Hz), 103.26 (d,
\emph{J} = 24.6 Hz).

\textsuperscript{19}F NMR (376 MHz, DMSO) ($\delta$): -60.40 (t, \emph{J} =
25.7 Hz, 2F, O-C\textbf{F\textsubscript{2}}-Ar), -111.61 (td, \emph{J} =
26.5, 13.0 Hz, 2F, Ar-\textbf{F}), -115.39 (t, \emph{J} = 11.0 Hz, 1F,
Ar-\textbf{F}), -132.89 (dd, \emph{J} = 22.0, 9.2 Hz, 2F,
Ar-\textbf{F}), -163.29 (tt, \emph{J} = 22.2, 6.0 Hz, 1F,
Ar-\textbf{F}).

\begin{figure}[H]
    \centering
    \includegraphics[width=0.3\linewidth]{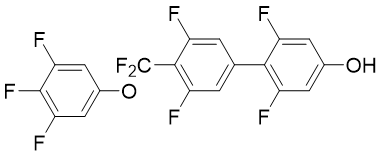}
\end{figure}

\sloppy\textbf{4-Hydroxy-5-(2,6-difluoro-4-hydroxyphenyl)-2-(difluoro(3,4,5-trifluorophenoxy)methyl)-
1,3-difluorobenzene}

Appearance: White powder

R\textsubscript{f} (DCM): 0.40

\textsuperscript{1}H NMR (400 MHz, DMSO) ($\delta$): 10.79 (s, 1H,
Ar-O\textbf{H}), 7.44 -- 7.35 (m, 4H, Ar-\textbf{H})*, 6.61 (ddd,
\emph{J} = 10.2, 2.8 Hz, 2H, Ar-\textbf{H}). (*Overlapping signals)

\textsuperscript{13}C\{\textsuperscript{1}H\} NMR (101 MHz, DMSO) ($\delta$):
161.00 (dd, \emph{J} = 246.2, 10.0 Hz), 160.41 -- 159.89
(m\textsubscript{apparent}), 157.49 (d, \emph{J} = 5.9 Hz), 150.27 (ddd,
\emph{J} = 248.9, 10.6, 5.3 Hz), 144.50 -- 143.68
(m\textsubscript{apparent}), 137.87 (dt, \emph{J} = 248.3, 15.3 Hz),
136.06 (t, \emph{J} = 11.9 Hz), 122.49, 119.86, 117.48 -- 116.99
(m\textsubscript{apparent}), 115.02, 114.75, 108.02, 107.78, 105.24 (t,
\emph{J} = 18.6 Hz), 99.83, 99.56.

\textsuperscript{19}F NMR (376 MHz, DMSO) ($\delta$): -60.71 (t, \emph{J} =
26.0 Hz, 2F, O-C\textbf{F\textsubscript{2}}-Ar), -111.79 (td, \emph{J} =
26.0, 11.7 Hz, 2F, Ar-\textbf{F}), -114.54 (d, \emph{J} = 10.8 Hz, 2F,
Ar-\textbf{F}), -132.92 (dd, \emph{J} = 22.0, 9.1 Hz, 2F,
Ar-\textbf{F}), -163.29 (tt, \emph{J} = 22.0, 6.1 Hz, 1F,
Ar-\textbf{F}).

\begin{figure}[H]
    \centering
    \includegraphics[width=0.3\linewidth]{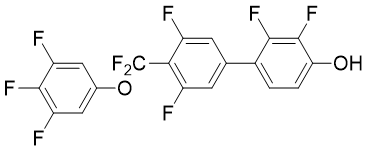}
\end{figure}

\sloppy\textbf{4-Hydroxy-5-(2,3-difluoro-4-hydroxyphenyl)-2-(difluoro(3,4,5-trifluorophenoxy)methyl)-
1,3-difluorobenzene}

Appearance: White powder

R\textsubscript{f} (DCM): 0.33

\textsuperscript{1}H NMR (400 MHz, DMSO) ($\delta$): 10.90 (s, 1H,
Ar-O\textbf{H}), 7.51 (d\textsubscript{apparent}, \emph{J} = 11.5 Hz,
2H, Ar-\textbf{H}), 7.47 -- 7.37 (m\textsubscript{apparent}, 2H,
Ar-\textbf{H}), 7.32 (td, \emph{J} = 8.7, 2.2 Hz, 1H, Ar-\textbf{H}),
6.90 (td, \emph{J} = 8.0, 1.9 Hz, 1H, Ar-\textbf{H}).

\textsuperscript{13}C NMR (101 MHz, DMSO) $\delta$ 159.48 (dd, \emph{J} =
255.2, 6.9 Hz), 151.89 (ddd, \emph{J} = 248.0, 10.8, 5.3 Hz), 150.15
(dd, \emph{J} = 239.6, 11.1 Hz), 148.45 (d, \emph{J} = 6.9 Hz), 144.59
(t, \emph{J} = 12.3 Hz), 141.54 (t, \emph{J} = 15.0 Hz), 139.20 (d,
\emph{J} = 14.4 Hz), 137.09 (t, \emph{J} = 15.7 Hz), 124.97 (t, \emph{J}
= 3.5 Hz), 116.89 (d, \emph{J} = 10.6 Hz), 113.81 (t, \emph{J} = 2.3
Hz), 113.39 (dt, \emph{J} = 24.2, 3.4 Hz), 108.40 (dd, \emph{J} = 23.8,
6.5 Hz), 107.53 (t, \emph{J} = 6.2 Hz).

\textsuperscript{19}F NMR (376 MHz, DMSO) ($\delta$): -60.43 (t, \emph{J} =
25.6 Hz, 2F, O-C\textbf{F\textsubscript{2}}-Ar ), -111.20 (td, \emph{J}
= 25.8, 11.9 Hz, 2F, Ar-\textbf{F}), -132.84 (dd, \emph{J} = 22.1, 9.3
Hz, 2F, Ar-\textbf{F}), -141.59 (dd, \emph{J} = 20.5, 8.4 Hz, 1F,
Ar-\textbf{F}), -160.90 (dd, \emph{J} = 20.6, 8.0 Hz, 1F,
Ar-\textbf{F}), -163.17 (tt, \emph{J} = 22.3, 5.4 Hz, 1F,
Ar-\textbf{F}).

\subsection*{3.2. General Esterification Protocol}

\begin{figure}[H]
    \centering
    \includegraphics[width=1\linewidth]{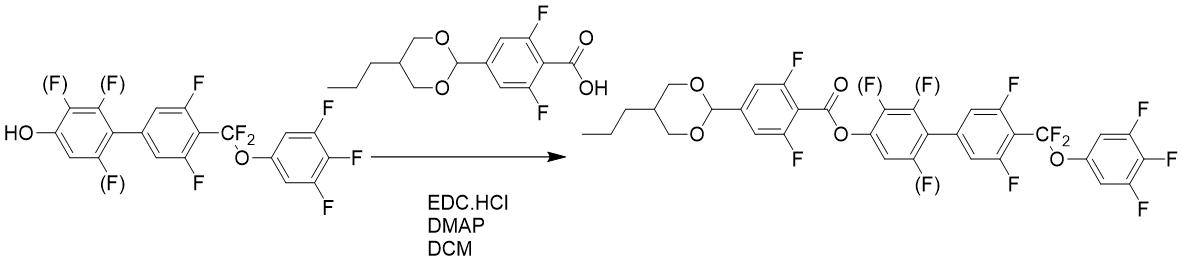}
\end{figure}

\textbf{Scheme 2.} General esterification protocol used in the
preparation of compounds \textbf{2-4}.

The 2,6-difluoro-4-(5-propyl-1,3-dioxan-2-yl)benzoic acid (1.5 mmol),
the relevant phenol (1 mmol), EDC.HCl (1.5 mmol), DMAP (5 mg) were
weighed into a reaction vial or round bottom flask. DCM was added
(\textasciitilde{} 10 ml, final concentration w.r.t. phenol
\textasciitilde{} 0.1 M), and the resulting suspension stirred until
complete consumption of the starting phenol as judged by TLC. The crude
reaction mixture was purified by flash chromatography using a 12 g SiO2
cartridge as the stationary phase and a gradient of hexane/DCM (0\% DCM
- 100\% DCM) as the mobile phase, with detection made in the range
200-800 nm. The chromatographed material was filtered through a 200 nm
PTFE syringe filter, concentrated to dryness, and finally recrystalised
from ethanol.

\begin{figure}[H]
    \centering
    \includegraphics[width=0.5\linewidth]{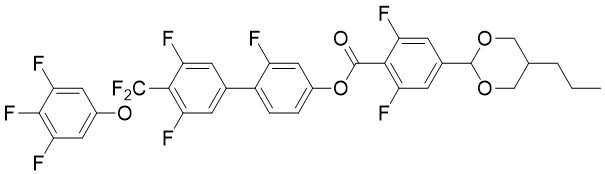}
\end{figure}

\textbf{2}

Yield: (white crystals) 516 mg, 75\%

R\textsubscript{f} (DCM): 0.91

\textsuperscript{1}H NMR (400 MHz, CDCl\textsubscript{3}) ($\delta$): 7.49 (t,
J = 8.6 Hz, 1H, Ar-\textbf{H}), 7.26 -- 7.16 (m, 6H, Ar-\textbf{H})*,
7.04 -- 6.95 (m\textsubscript{apparent}, 2H, Ar-\textbf{H}), 5.40 (s,
1H, Ar-C\textbf{H}-O\textsubscript{2}), 4.26 (dd, J = 11.8, 4.7 Hz, 2H,
O-C\textbf{H}\textsubscript{ax}(H\textsubscript{eq})-CH), 3.54 (t, J =
11.2 Hz, 2H, O-C\textbf{H}\textsubscript{eq}(H\textsubscript{ax})-CH),
2.21 -- 2.08 (m\textsubscript{apparent}, 1H,
(CH\textsubscript{2})\textsubscript{2}-C\textbf{H}-CH\textsubscript{2}),
1.35 (h, J = 7.6 Hz, 2H,
CH\textsubscript{2}-C\textbf{H\textsubscript{2}}-CH\textsubscript{3}),
1.11 (q, J = 7.6 Hz, 2H,
CH-C\textbf{H\textsubscript{2}}-CH\textsubscript{2}-CH\textsubscript{3}
), 0.94 (t, J = 7.3 Hz, 3H,
CH\textsubscript{2}-C\textbf{H\textsubscript{3}}). (*Overlapping
Signals)

\textsuperscript{13}C\{\textsuperscript{1}H\} NMR (101 MHz,
CDCl\textsubscript{3}) ($\delta$): 162.21 (dd, \emph{J} = 248.9, 6.1 Hz),
161.27 (d\textsubscript{apparent}, \emph{J} = 260.9 Hz), 159.56 (d,
\emph{J} = 253.7 Hz), 151.58 (d, \emph{J} = 11.8 Hz), 151.03 (dt,
\emph{J} = 259.7, 10.9 Hz), 145.74 (t, \emph{J} = 9.5 Hz), 144.59 (t,
\emph{J} = 14.8 Hz), 140.67 (t, \emph{J} = 12.9 Hz), 130.62 (d, \emph{J}
= 3.7 Hz), 123.71 (d, \emph{J} = 13.6 Hz), 120.15, 118.35 (d, \emph{J} =
3.8 Hz), 113.14 (dt, \emph{J} = 24.4, 3.1 Hz), 110.84 (d, \emph{J} =
26.0 Hz), 110.33 (dd, \emph{J} = 23.9, 2.8 Hz), 109.33 (d, \emph{J} =
16.2 Hz), 107.48 (m\textsubscript{apparent}), 98.79 (t, \emph{J} = 2.2
Hz), 72.60, 33.90, 30.23, 19.53, 14.19.

\textsuperscript{19}F NMR (376 MHz, CDCl\textsubscript{3}) ($\delta$): -61.78
(t, \emph{J} = 26.3 Hz, 2F, O-C\textbf{F\textsubscript{2}}-Ar), -108.41
(d, \emph{J} = 9.6 Hz, 2F, Ar-\textbf{F}), -110.33 (td, \emph{J} = 26.4,
11.1 Hz, 2F, Ar-\textbf{F}), -113.60 (t, \emph{J} = 9.6 Hz, 1F,
Ar-\textbf{F}), -132.43 (dd, \emph{J} = 20.6, 8.7 Hz, 2F,
Ar-\textbf{F}), -163.11 (tt, \emph{J} = 20.8, 5.6 Hz, 1F, Ar-\textbf{F})

HPLC (C18): 99.68 \% (254 nm, 6.047 min)

HRMS (ESI +, m/z) = {[}m+Na{]}\textsuperscript{+}: Calculated for
C\textsubscript{33}H\textsubscript{22}F\textsubscript{10}O\textsubscript{5}Na:
711.1200, found 711.1182 (error = 2.5 ppm)

\begin{figure}
    \centering
    \includegraphics[width=0.5\linewidth]{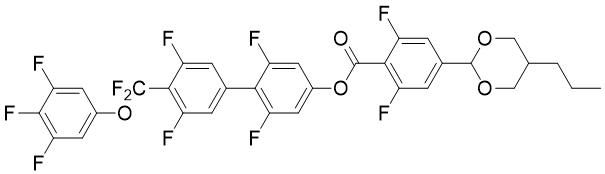}
\end{figure}

\textbf{3}

Yield: (white crystals) 502 mg, 71 \%

R\textsubscript{f} (DCM): 0.88

\textsuperscript{1}H NMR (400 MHz, CDCl\textsubscript{3}) ($\delta$): 7.20
(d\textsubscript{apparent}, J = 10.1 Hz, 2H, Ar-\textbf{H}), 7.16
(d\textsubscript{apparent}, J = 10.4 Hz, 2H, Ar-\textbf{H}), 7.09 --
6.94 (m, 4H, Ar-\textbf{H})*, 5.40 (s, 1H,
Ar-C\textbf{H}-O\textsubscript{2}), 4.26 (dd, J = 11.8, 4.6 Hz, 2H,
O-C\textbf{H}\textsubscript{ax}(H\textsubscript{eq})-CH), 3.54 (t, J =
11.3 Hz, 2H, O-C\textbf{H}\textsubscript{eq}(H\textsubscript{ax})-CH)),
2.21 -- 2.07 (m\textsubscript{apparent}, 1H,
(CH\textsubscript{2})\textsubscript{2}-C\textbf{H}-CH\textsubscript{2}),
1.35 (h, J = 7.4 Hz, 2H,
CH\textsubscript{2}-C\textbf{H\textsubscript{2}}-CH\textsubscript{3}),
1.11 (q, J = 7.0 Hz, 2H,
CH-C\textbf{H\textsubscript{2}}-CH\textsubscript{2}-CH\textsubscript{3}),
0.94 (t, J = 7.3 Hz, 3H,
CH\textsubscript{2}-C\textbf{H\textsubscript{3}}). (*Overlapping
Signals)

\textsuperscript{13}C\{\textsuperscript{1}H\} NMR (101 MHz,
CDCl\textsubscript{3}) ($\delta$): 162.32 (dd, J = 258.5, 5.7 Hz), 161.00,
160.23 (dd, J = 251.6, 7.8 Hz), 158.79, 152.36 (ddd, J = 251.1, 10.3,
5.6 Hz), 151.38 (t, J = 14.3 Hz), 146.05 (t, J = 9.7 Hz), 139.76 (t, J =
14.0 Hz), 134.28 (t, J = 10.0 Hz), 122.68, 120.02, 117.38, 114.76 (d, J
= 24.7 Hz), 110.36 (dd, J = 23.6, 3.5 Hz), 108.91, 107.75 -- 107.29
(m\textsubscript{apparernt}), 106.97 -- 106.47
(m\textsubscript{apparent}), 98.73 (t, J = 2.0 Hz), 72.60, 33.90, 30.22,
19.53, 14.18.

\textsuperscript{19}F NMR (376 MHz, CDCl\textsubscript{3}) ($\delta$): -61.95
(t, \emph{J} = 26.5 Hz, 2F, O-C\textbf{F\textsubscript{2}}-Ar), -108.15
(d, \emph{J} = 10.0 Hz, 2F, Ar-\textbf{F}), -110.53 (td, \emph{J} =
26.5, 10.7 Hz, 2F, Ar-\textbf{F}), -111.71 (d, \emph{J} = 9.1 Hz, 2F,
Ar-\textbf{F}), -132.43 (dd, \emph{J} = 20.9, 8.6 Hz, 2F,
Ar-\textbf{F}), -163.07 (tt, \emph{J} = 20.7, 5.9 Hz, 1F,
Ar-\textbf{F}).

HPLC (C18): 99.58 \% (254 nm, 6.060 min)

HRMS (ESI +, m/z) = {[}m+Na{]}\textsuperscript{+}: Calculated for
C\textsubscript{33}H\textsubscript{21}F\textsubscript{11}O\textsubscript{5}Na:
729.1106, found 729.1101 (error = 0.7 ppm)

\begin{figure}[H]
    \centering
    \includegraphics[width=0.5\linewidth]{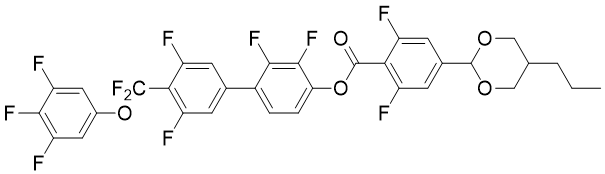}
\end{figure}

\textbf{4}

Yield: (white crystals) 435 mg, 62 \%

R\textsubscript{f} (DCM): 0.90

\textsuperscript{1}H NMR (400 MHz, CDCl\textsubscript{3}) ($\delta$): 7.29 --
7.16 (m, 6H, Ar-\textbf{H})*\textsuperscript{†}, 7.00 (dd, \emph{J} =
7.9, 5.8 Hz, 2H, Ar-\textbf{H}), 5.41 (s, 1H,
Ar-C\textbf{H}-O\textsubscript{2}), 4.26 (dd, \emph{J} = 11.7, 4.6 Hz,
2H, O-C\textbf{H}\textsubscript{ax}(H\textsubscript{eq})-CH), 3.54 (t,
\emph{J} = 11.4 Hz, 2H,
O-C\textbf{H}\textsubscript{eq}(H\textsubscript{ax})-CH)), 2.21 -- 2.07
(m\textsubscript{apparent}, 1H,
(CH\textsubscript{2})\textsubscript{2}-C\textbf{H}-CH\textsubscript{2}),
1.35 (h, \emph{J} = 7.7 Hz, 2H,
CH\textsubscript{2}-C\textbf{H\textsubscript{2}}-CH\textsubscript{3}),
1.11 (q, \emph{J} = 7.1 Hz, 2H,
CH-C\textbf{H\textsubscript{2}}-CH\textsubscript{2}-CH\textsubscript{3}),
0.94 (t, \emph{J} = 7.3 Hz, 3H,
CH\textsubscript{2}-C\textbf{H\textsubscript{3}}). (*Overlapping peaks,
\textsuperscript{†}Overlapping CDCl\textsubscript{3} peak)

\textsuperscript{13}C\{\textsuperscript{1}H\} NMR (101 MHz,
CDCl\textsubscript{3}) ($\delta$): 162.43 (dd, \emph{J} = 259.8, 5.4 Hz),
161.34 (dd, \emph{J} = 259.2, 6.3 Hz), 158.22, 150.82 (ddd, \emph{J} =
250.5, 10.6, 5.1 Hz), 150.03, 147.45 (dd, \emph{J} = 245.7, 12.0 Hz),
146.18 (t, \emph{J} = 9.9 Hz), 144.85 -- 144.33
(m\textsubscript{apparent}), 142.49 (d, \emph{J} = 15.0 Hz), 140.04 --
139.30 (m\textsubscript{apparent}), 137.35 (t, \emph{J} = 15.1 Hz),
125.49 (d, \emph{J} = 9.5 Hz), 123.73 (t, \emph{J} = 4.0 Hz), 122.69,
120.04, 119.15 (d, \emph{J} = 4.0 Hz), 113.14 (dt, \emph{J} = 24.7, 3.3
Hz), 110.38 (dd, \emph{J} = 23.5, 3.5 Hz), 109.58 (t, \emph{J} = 14.3
Hz), 108.43 (t, \emph{J} = 16.8 Hz), 107.80 -- 107.30
(m\textsubscript{apparent}), 98.73 (t, \emph{J} = 1.6 Hz), 72.58, 33.90,
30.22, 19.53, 14.16.

\textsuperscript{19}F NMR (376 MHz, CDCl\textsubscript{3}) ($\delta$): -61.87
(t, \emph{J} = 26.4 Hz, 2F, O-C\textbf{F\textsubscript{2}}-Ar), -107.59
(d, \emph{J} = 10.0 Hz, 2F, Ar-\textbf{F}), -109.76 (td, \emph{J} =
26.5, 10.7 Hz, 2F, Ar-\textbf{F}), -132.40 (dd, \emph{J} = 20.7, 8.7 Hz,
2F, Ar-\textbf{F}), -138.63 (dd, \emph{J} = 20.1, 6.5 Hz, 1F,
Ar-\textbf{F}), -147.62 (dd, \emph{J} = 20.3, 5.6 Hz, 1F,
Ar-\textbf{F}), -163.04 (tt, \emph{J} = 21.0, 5.0 Hz, 1F,
Ar-\textbf{F}).

HPLC (C18): 99.81 \% (254 nm, 6.061 min)

HRMS (ESI +, m/z) = {[}m+Na{]}\textsuperscript{+}: Calculated for
C\textsubscript{33}H\textsubscript{21}F\textsubscript{11}O\textsubscript{5}Na:
729.1106, found 729.1101 (error = 0.6 ppm)

\subsection*{3.3. Example NMR Spectra, HPLC Chromatograms, and HRMS Spectra.}

\includegraphics[width=5.7874in,height=4.03823in]{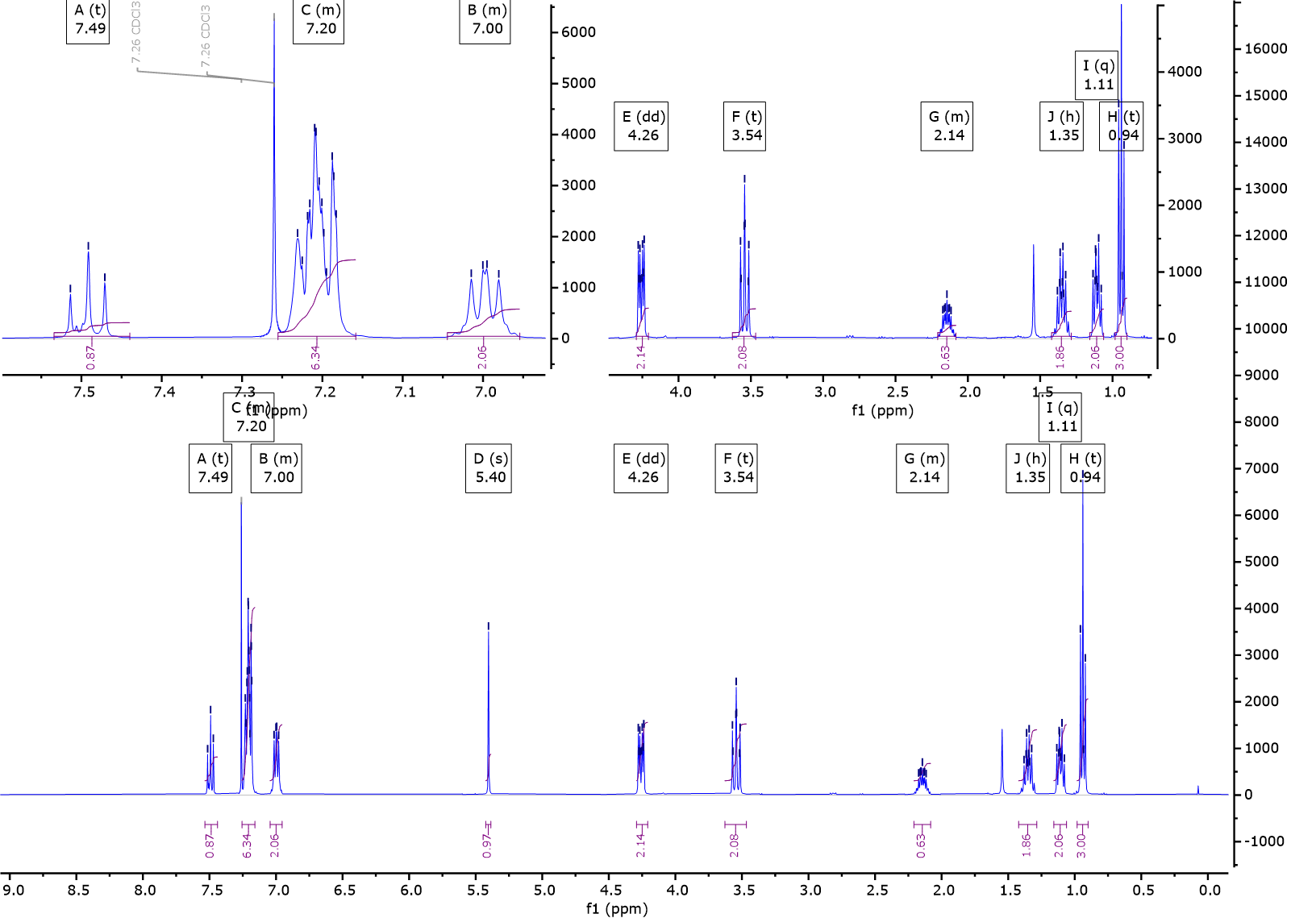}

\includegraphics[width=5.7874in,height=4.05939in]{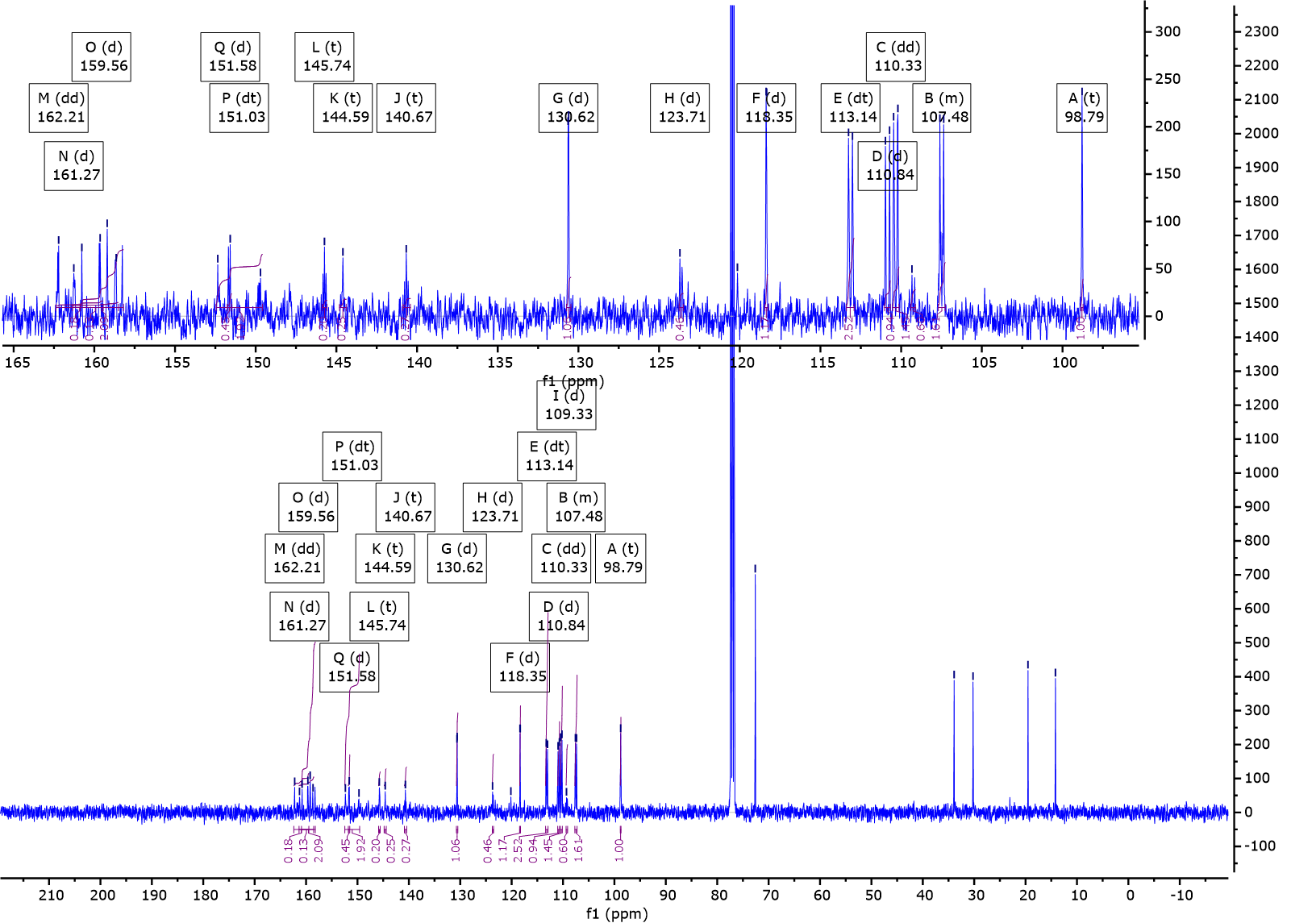}

\includegraphics[width=5.7874in,height=4.03823in]{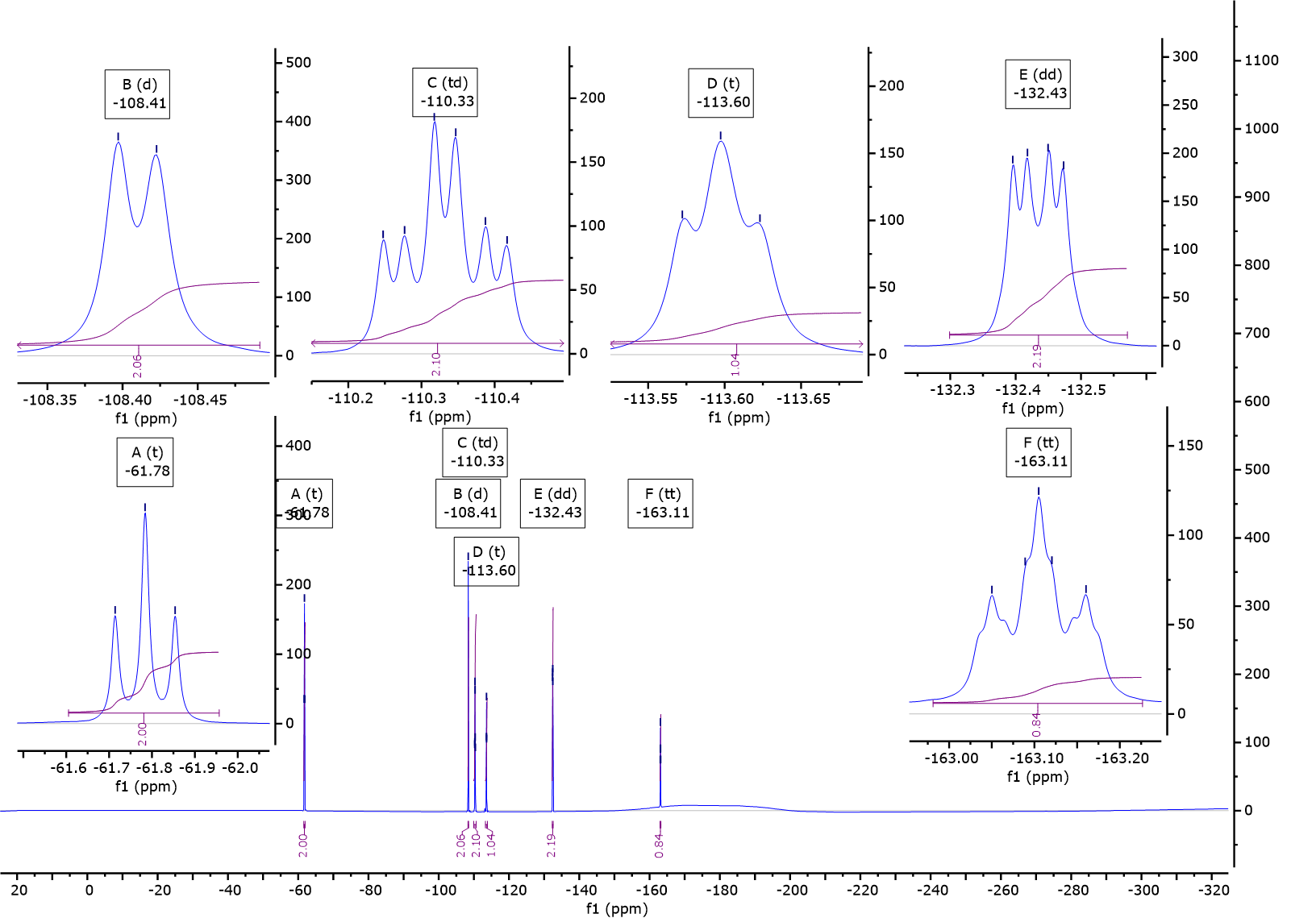}

\includegraphics[width=6.26806in,height=2.78403in]{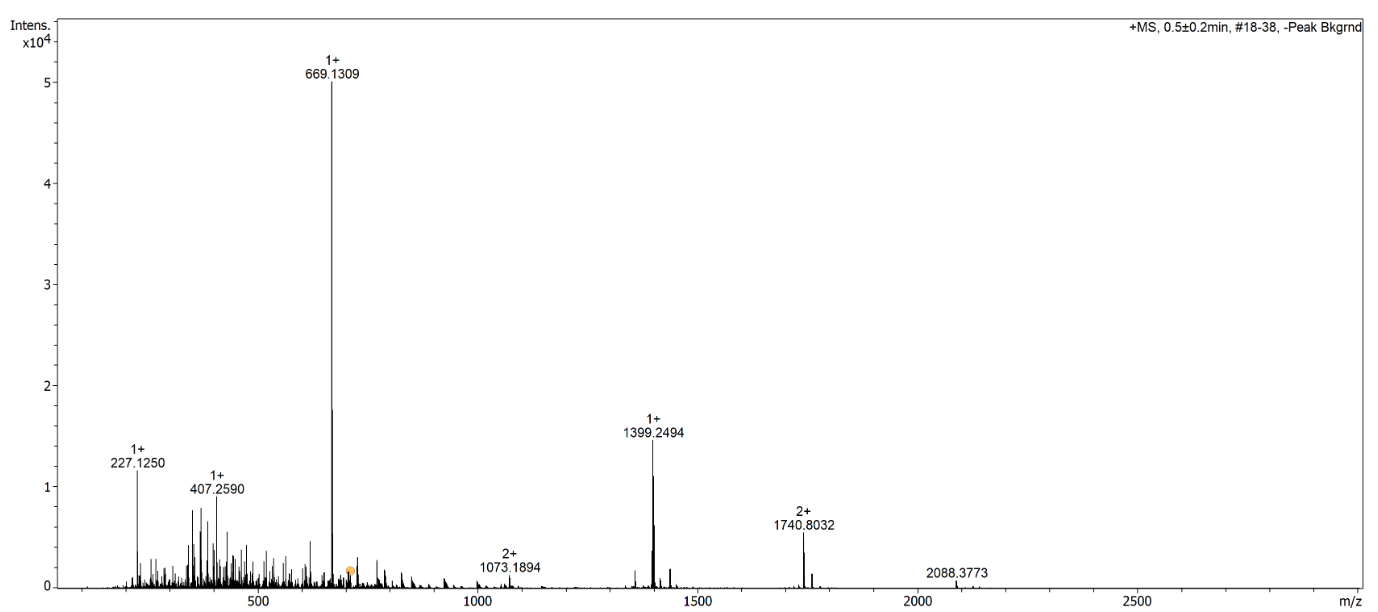}

Fig. S14. \textsuperscript{1}H {[}top{]},
\textsuperscript{13}C\{\textsuperscript{1}H\} {[}middle{]}, and
\textsuperscript{19}F {[}bottom{]} NMR spectra, and HRMS Spectra for
\textbf{2}.

\includegraphics[width=5.7874in,height=4.03823in]{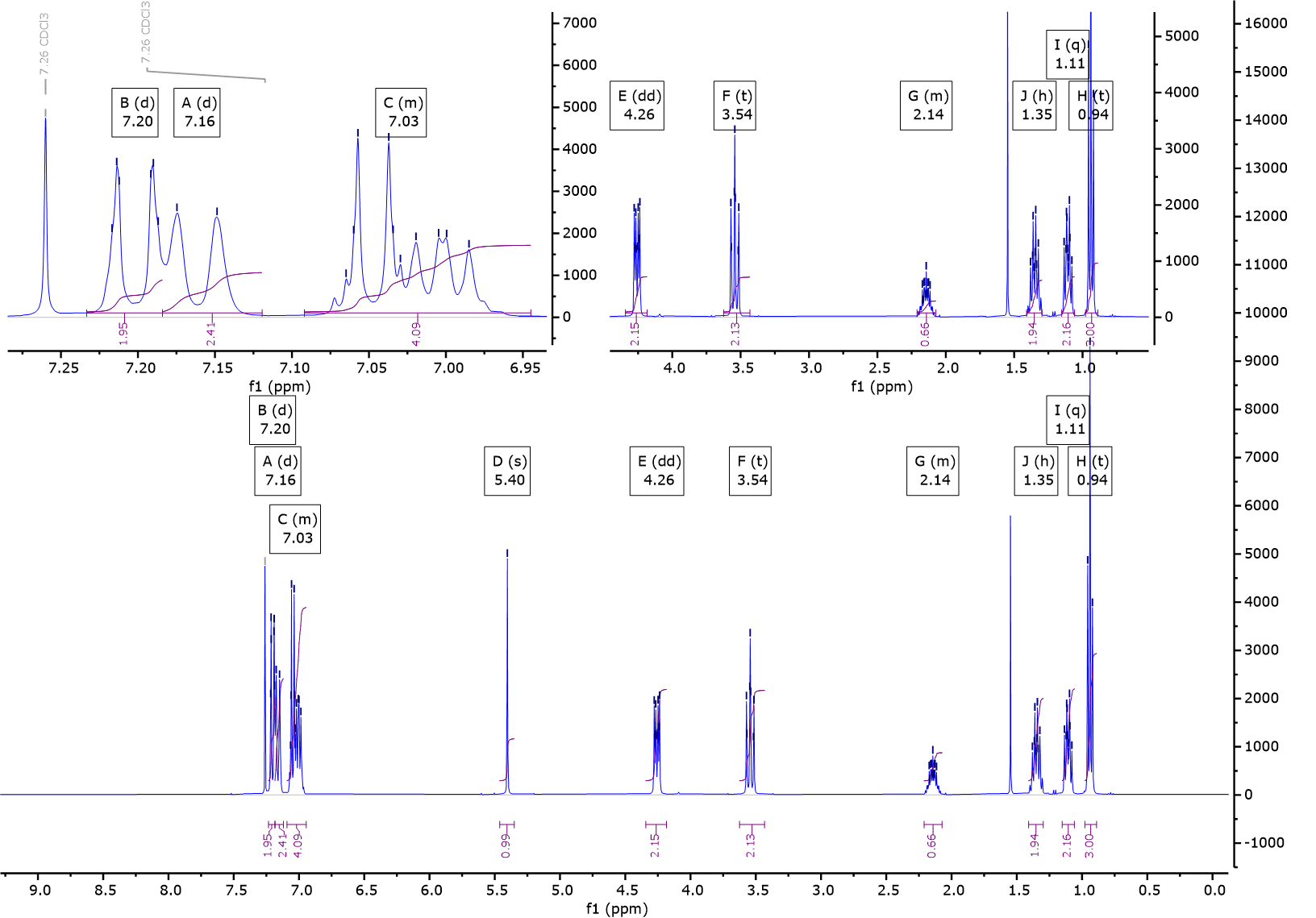}

\includegraphics[width=5.7874in,height=4.0549in]{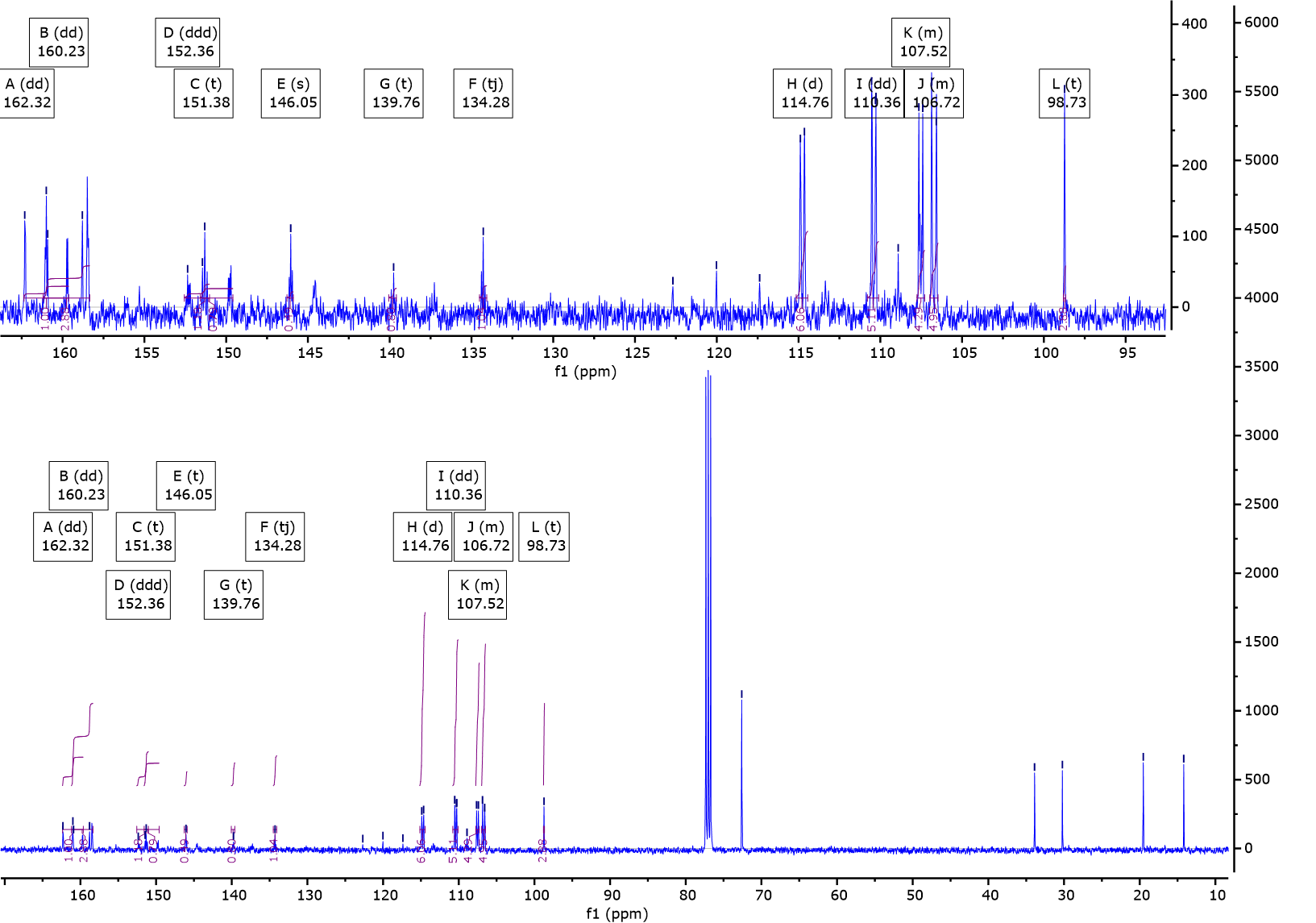}

\includegraphics[width=5.7874in,height=4.03823in]{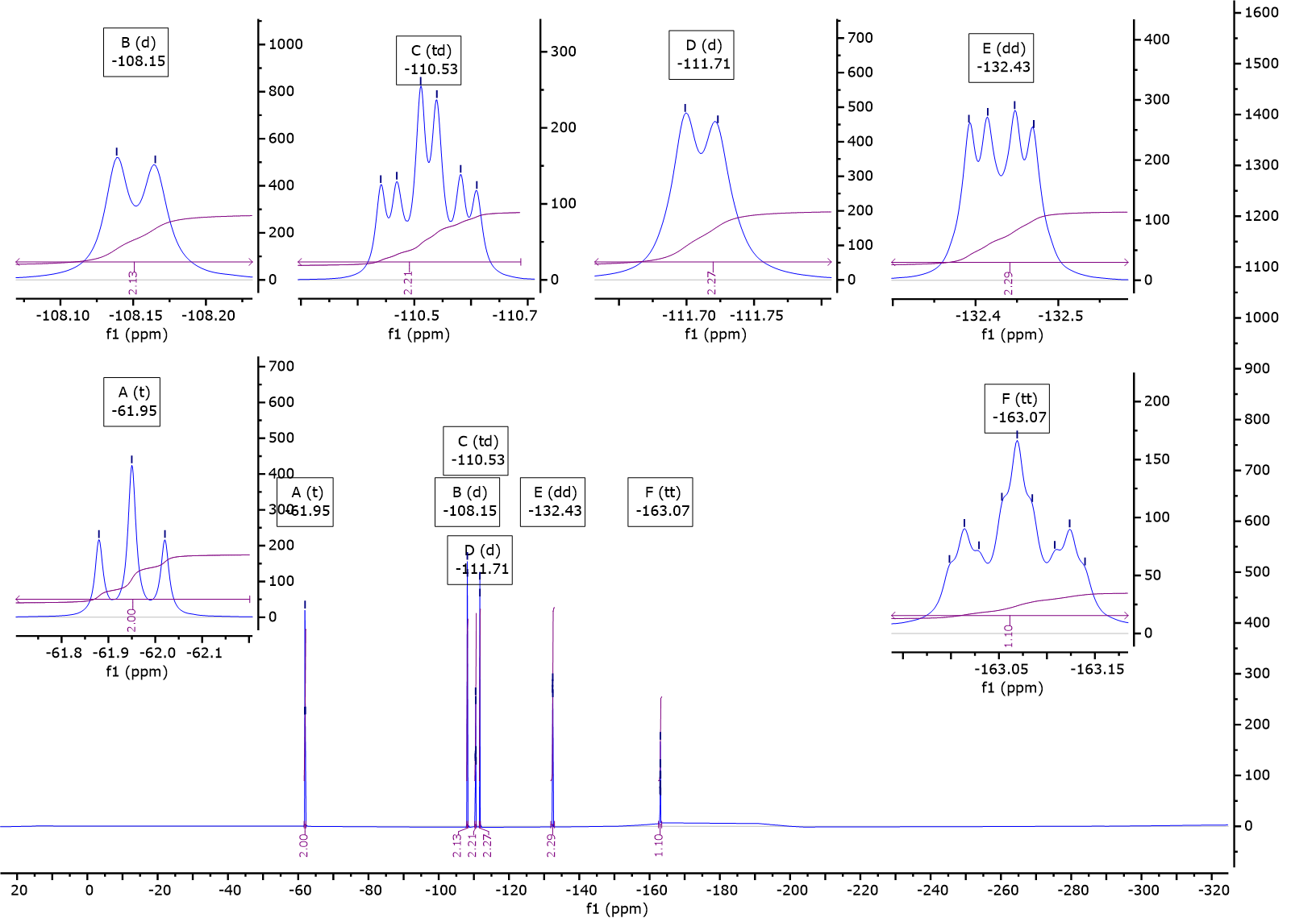}

\includegraphics[width=6.26772in,height=2.8128in]{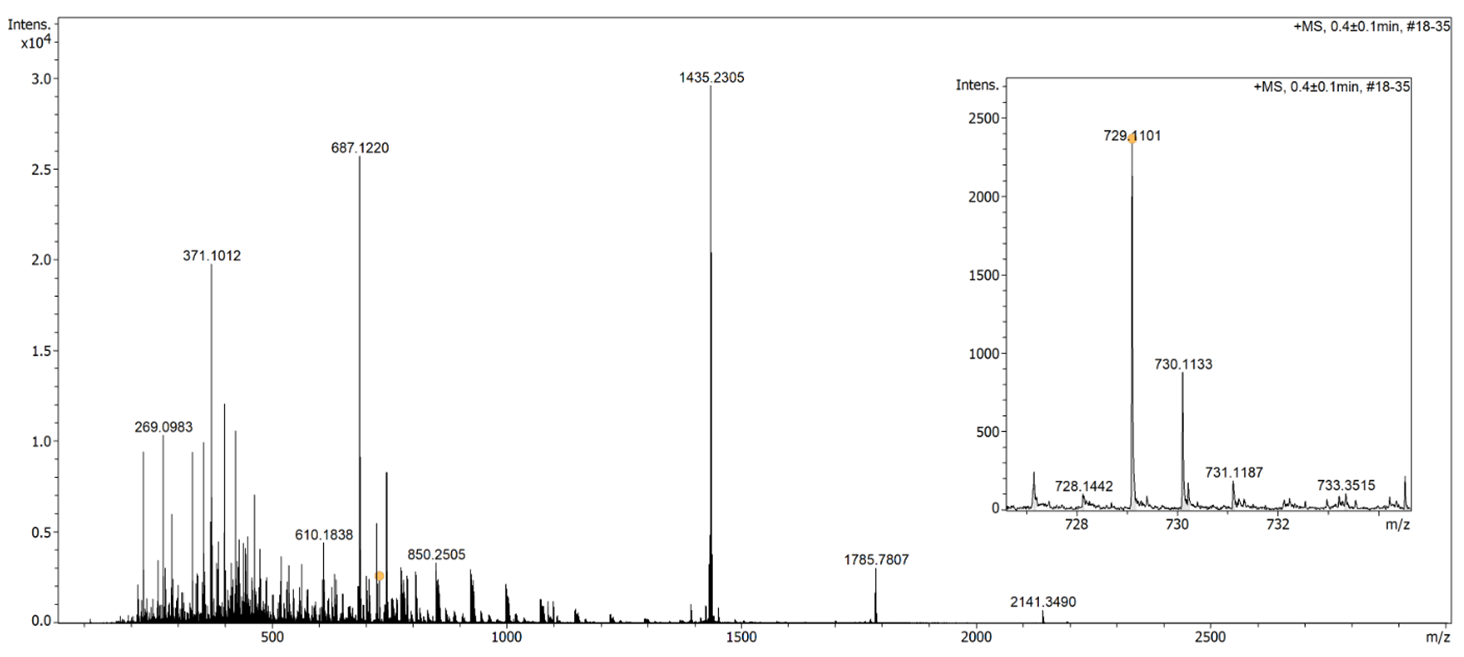}

Fig. S15. \textsuperscript{1}H {[}top{]},
\textsuperscript{13}C\{\textsuperscript{1}H\} {[}middle{]}, and
\textsuperscript{19}F {[}bottom{]} NMR spectra, and HRMS Spectra for
\textbf{3}.

\includegraphics[width=5.7874in,height=4.03823in]{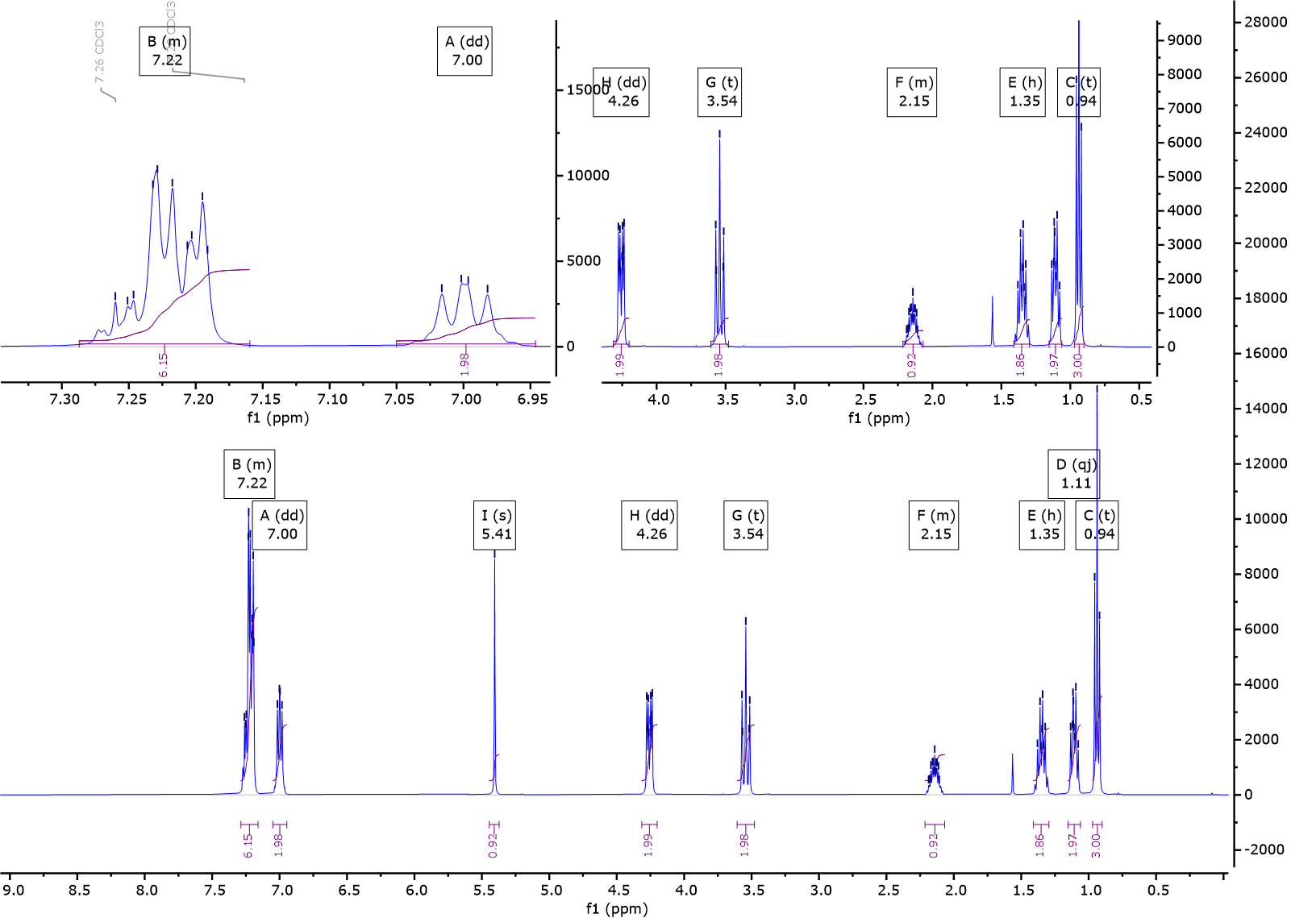}

\includegraphics[width=5.7874in,height=4.03823in]{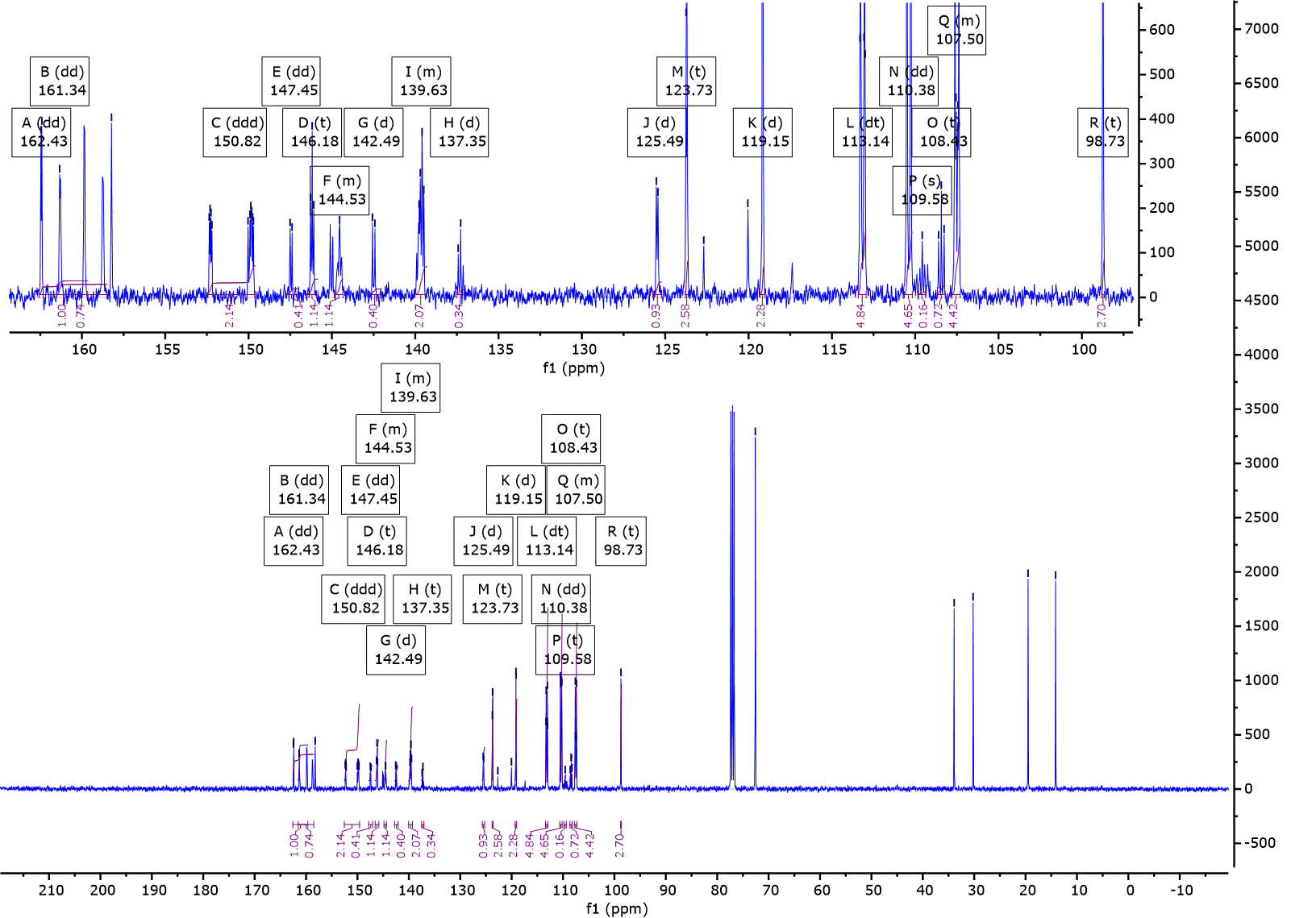}

\includegraphics[width=5.7874in,height=3.9459in]{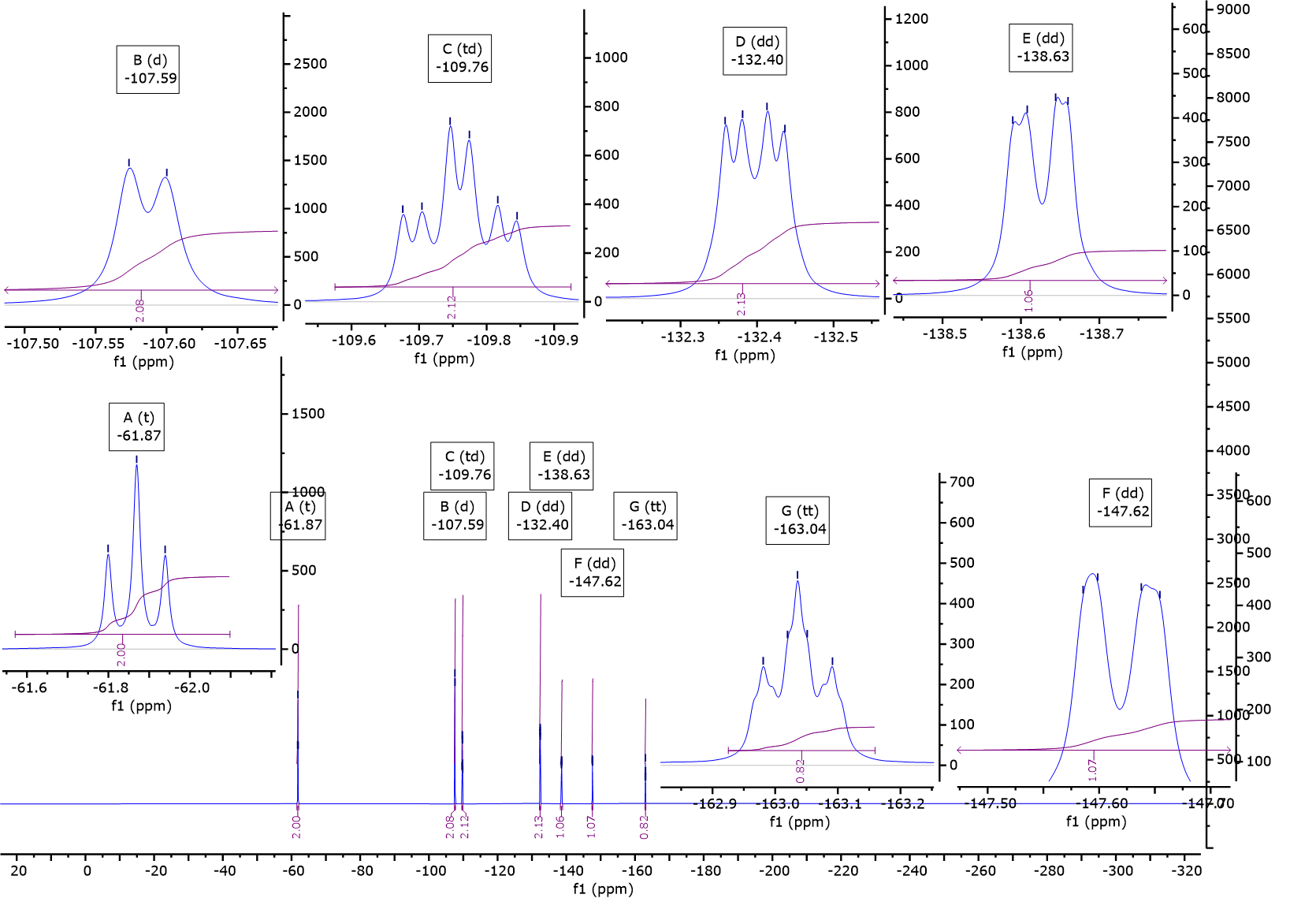}

\includegraphics[width=6.26772in,height=2.78141in]{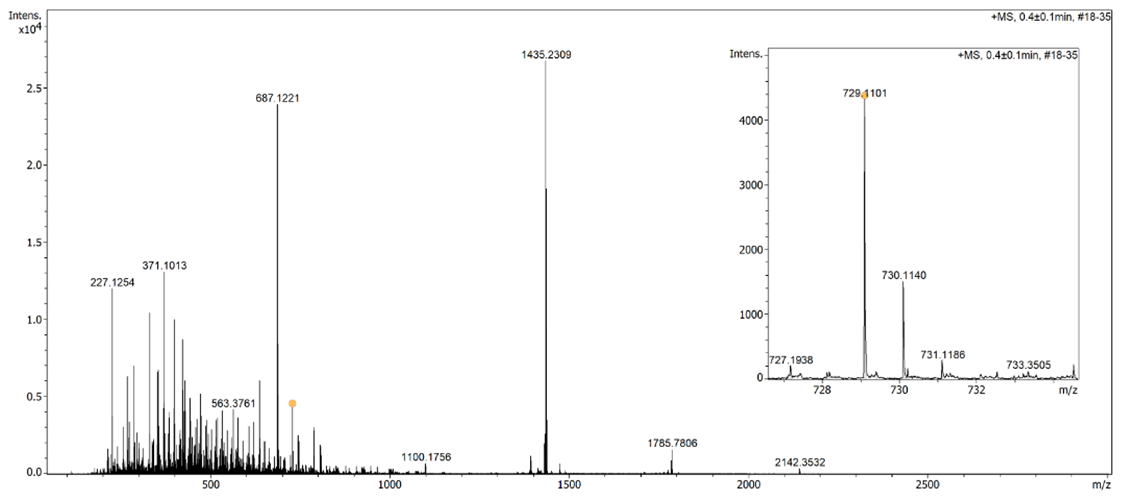}

Fig. S16. \textsuperscript{1}H {[}top{]},
\textsuperscript{13}C\{\textsuperscript{1}H\} {[}middle{]}, and
\textsuperscript{19}F {[}bottom{]} NMR spectra, and HRMS Spectra for
\textbf{4}.

\section{Supplemental references}

{[}1{]} Gibb, C.J., Hobbs, J., Nikolova, D.I. et al. Spontaneous
symmetry breaking in polar fluids. Nat Commun, 15, 5845 (2024)

{[}2{]} Hobbs, J., Gibb, C.J. and Mandle, R.J. Emergent
Antiferroelectric Ordering and the Coupling of Liquid Crystalline and
Polar Order. Small Sci., 4, 2400189 (2024)